\documentclass[prx,reprint,superscriptaddress,amsmath,amssymb,aps,longbibliography]{revtex4-2}

\usepackage{natbib}
\usepackage{amsmath}
\usepackage{amssymb}
\usepackage{amsthm}
\usepackage{amsfonts}
\usepackage[caption=false]{subfig}
\usepackage[colorlinks]{hyperref}
\hypersetup{
    citecolor = blue,
    urlcolor = blue
}
\usepackage[all]{hypcap}
\usepackage{tikz}
\usepackage{relsize}
\usepackage{color,soul}
\usepackage[utf8]{inputenc}
\usepackage{capt-of}
\usepackage{mathtools}
\usepackage{float}
\usepackage[section]{placeins}
\usepackage{listings}
\usepackage{svg}
\usepackage{xifthen}
\usepackage{multirow}
\usepackage{makecell}
\usepackage{siunitx}
\usepackage{hyperref}
\usepackage[normalem]{ulem}
\usepackage{bm}
\usepackage{esint}

\newcommand{\google}{\affiliation{Google Quantum AI, Santa Barbara, CA 93117, USA}}
\newcommand{\equal}{\thanks{These authors contributed equally: vladkurilovich@google.com, gabriellelcr@google.com, opremcak@google.com}}
\newcommand{\ket}[1]{{|#1\rangle}}

\makeatother

\graphicspath{{figures/}}

\begin{document}
\widetext

\title{
Correlated Error Bursts in a Gap-Engineered Superconducting Qubit Array
}

\author{Vladislav D. Kurilovich}
\equal
\author{Gabrielle Roberts}
\equal
\author{Leigh S.~Martin}
\author{Matt McEwen}
\author{Alec Eickbusch}
\author{Lara Faoro}
\author{Lev B.~Ioffe}
\author{Juan Atalaya}
\author{Alexander Bilmes}
\author{John Mark Kreikebaum}
\author{Andreas Bengtsson}
\author{Paul Klimov}
\author{Matthew Neeley}
\author{Wojciech Mruczkiewicz}
\author{Kevin Miao}
\author{Igor L.~Aleiner}
\author{Julian Kelly}
\author{Yu Chen}
\author{Kevin Satzinger}
\author{Alex Opremcak}
\equal
\google

\date{\today}
\begin{abstract}
One of the roadblocks towards the implementation of a fault-tolerant superconducting quantum processor is {impacts} of ionizing radiation with the qubit substrate. Such impacts temporarily elevate the density of quasiparticles (QPs) across the device, leading to correlated qubit error bursts. The most damaging errors---$T_1$ errors---stem from QP tunneling across the qubit Josephson junctions (JJs). Recently, we demonstrated [\href{https://journals.aps.org/prl/pdf/10.1103/PhysRevLett.133.240601}{Phys.~Rev.~Lett.~{\bf 133}, 240601 (2024)}] that this type of error can be strongly suppressed by engineering the profile of superconducting gap at the {JJs} in a way that prevents QP tunneling. In this work, we identify a new type of impact-induced correlated error that persists in the presence of gap engineering. We observe that impacts shift the frequencies of the affected qubits, and thus lead to correlated {\it phase} errors. The frequency shifts are systematically negative, reach values up to $3\,{\rm MHz}$, and last for $\sim 1\,{\rm ms}$.
We provide evidence that the shifts originate from QP-qubit interactions in the JJ region.
Further, we demonstrate that {the shift-induced} phase errors can be detrimental to the performance of quantum error correction protocols.
\end{abstract}

\maketitle

\section{Introduction}\label{sec:introduction}

\begin{figure}[t]
  \begin{center}
    \includegraphics[scale = 1]{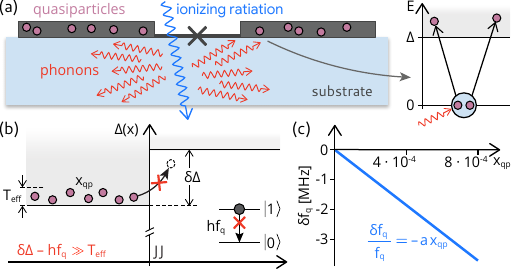}
    \caption{\textbf{The fallout from ionizing radiation in gap engineered qubits.} \textbf{(a)} Ionizing radiation impacts deposit large energies in the qubit substrate, which propagate throughout the device in the form of high-energy phonons. These phonons break Cooper pairs in the superconductors leading to elevated  {quasiparticle (QP)} densities, $x_\text{qp}$. The presence of QPs near the  {Josephson junctions (JJs)} of the qubits results in correlated errors. \textbf{(b)}  {A difference $\delta \Delta$ in the superconducting gaps of the JJ leads} exponentially suppresses QP tunneling, provided the characteristic QP energy, $T_\text{eff}$, satisfies $\delta \Delta - h f_{\rm q} \gg T_\text{eff}$ [$f_{\rm q}$ is the qubit frequency]. This protects the qubits from QP-induced $T_1$ errors. \textbf{(c)} Even though QPs cannot tunnel, their elevated density results in qubit frequency shifts $\delta f_{\rm q}$ and thus leads to correlated phase errors  {[to produce the plot, we use Eq.~\eqref{eq:shift} with $a = 0.77$ computed in Appendix~\ref{appendix:shifts}]}.
    } 
    \label{fig:fig1}
  \end{center}
\end{figure}

Realization of a fault-tolerant quantum processor hinges on the ability to correct errors affecting individual physical qubits.
When errors are rare and uncorrelated, they can be dealt with using quantum error correction (QEC).
The performance of {many} QEC codes (e.g., the surface code) is expected to improve exponentially with the number of physical qubits; in principle, one should be able to reach an arbitrarily low logical error rate (LER) by scaling up the system \cite{fowler_2012}.
However, this scaling behavior breaks down when correlated error \textit{bursts} exist that simultaneously affect a large fraction of qubits.
Such bursts then set a floor on the LER for QEC codes.

In superconducting quantum processors,  {one source of correlated error bursts} is \textit{impacts} of ionizing radiation with the qubit substrate \cite{vepsalainen_2020, wilen_2021, cardani_2021, martinis2021, mcewen_2022, li_2025, valenti_2025, larson_2025, nho_2025}. 
The impacts generate a large number of high-energy phonons in the substrate that rapidly spread across the qubit array, see Fig.~\ref{fig:fig1}(a).
 {The superconducting films} comprising the qubits efficiently absorb these phonons; 
in the process, the phonon energy converts into Bogoliubov quasiparticle (QP) excitations.
It is the proliferation of QPs that leads to correlated qubit errors. 
The most detrimental situation is realized when QPs tunnel across the qubits' Josephson junctions (JJs). 
A tunneling QP strongly couples to the electric field of the qubit, and readily absorbs a qubit excitation \cite{lenander_2011, catelani_relaxation_2011, catelani_quasiparticle_2011, riste_2013, serniak_2018, serniak_2019, diamond_2022}.
 {As a result,} the qubit $T_1$ degrades to sub-${\rm \mu s}$ scale following impacts \cite{mcewen_2022}.
The degradation persists for thousands of QEC cycles (a cycle duration is $\sim 1\,{\rm \mu s}$ \cite{google_2022, krinner2022, google_2024}) and, therefore, leads to a logical error with a high probability~\cite{chen_2021}.

Fortunately, there exists a way to inhibit QP tunneling at the hardware level. The idea is to engineer the profile of the superconducting gap across the JJs to form a potential barrier for QPs, see Fig.~\ref{fig:fig1}(b). 
If the barrier is sufficiently large then QPs cannot tunnel
\cite{sun_2012, wang_2014, kalashnikov_2020, marchegiani_2022, kamenov_2023, connolly_2024, harrington_2024}.
This protects the qubits from QP-induced $T_1$ errors.
In an earlier work, we verified the effectiveness of this strategy in suppressing correlated $T_1$ error bursts \cite{mcewen_2024}.

While gap engineering dramatically reduces the severity of error bursts, it {\it does not} remove the bursts entirely. 
{Indeed, recent work~\cite{google_2024} has shown that the LER floor of a repetition code in a gap-engineered qubit array remains set by error bursts.}
These ``residual'' bursts are less frequent than before, now occurring on hour rather than second timescales, and have a shorter recovery time ($\sim 1\,{\rm ms}$ vs.~$\sim 25\,{\rm ms}$). 
What mechanism causes them, and why {does it persist with gap engineering}?
The QEC data of Ref.~\cite{google_2024} offers little insight into these questions. 
Yet, answering them is crucial for the development of a fault-tolerant superconducting quantum processor.

To understand the origin of error bursts in gap-engineered qubit arrays, we perform a rapid, repetitive sampling experiment that simultaneously monitors the time evolution of coherence across all qubits in the array. 
The key innovation in comparison to previous experiments is that we supplement measurements of qubit energy relaxation {with} Ramsey and spin-echo sequences. 
We find that, rather than $T_1$ errors, the main signature of radiation impacts in gap-engineered arrays are quasi-static qubit frequency shifts, and correlated {\it phase errors} that result from them.
The frequency shifts are systematically negative, have magnitude in MHz range,  {and {recover} over $\sim 1\,{\rm ms}$ timescale following impacts}. 
The shape of the recovery points towards the QP-based origin of the shifts:
it is compatible with the kinetics of elevated QP density in the JJ region and is governed by recombination processes. 
We explain theoretically how QPs shift the qubit frequencies {in the absence of tunneling}, and relate the frequency shift to QP density [cf.~Fig.~\ref{fig:fig1}(c)].

The characteristic duration of a QEC cycle is $\approx 1\,{\rm \mu s}$~\cite{google_2024}.
In this time, a frequency shift of $1\,{\rm MHz}$ leads to a spurious phase accumulation of $\approx 2\pi$.
If this phase is not accurately compensated for in the QEC circuit, 
it may lead to logical errors in QEC experiments.
To directly correlate the QEC bursts with phase errors, we run coherence characterization in parallel with a repetition code on two adjacent parts of the qubit array.
We observe a strong spatiotemporal correlation between phase errors on monitor qubits and {error detection events} in the repetition code.

In addition to frequency shifts, we observe that each impact event begins with a transient $T_1${-}error burst lasting on the order of $10\,{\rm \mu s}$.
On the one hand, the duration of the $T_1$ error bursts on the gap engineered device is shorter than {on} devices without gap engineering by more than two orders of magnitude.
On the other hand, it still exceeds the typical duration of QEC cycle $\sim 1\,{\rm \mu s}$.
Together with the observation of large frequency shifts, this highlights the need for further {mitigation} strategies {of} radiation impacts, e.g., using phonon- or quasiparticle-traps \cite{riwar_2016, riwar_2019, iaia_2022, yelton_2024}.

\section{Experimental device}

{Our Willow processor} features an array of 72 superconducting qubits patterned on a silicon substrate, of which we utilize a subset of 60 qubits in our experiments {\cite{google_2024}}. 
The qubits are {aluminum-based} frequency tunable transmons, with operating frequencies in the interval $5.9$--$6.5$~GHz.
Each qubit in the array can be individually controlled,  measured, and reset, see Appendix~\ref{appendix:device} and Ref.~\cite{google_2024} for details.

Protection from ionizing radiation is provided by superconducting gap engineering \cite{mcewen_2024}: the qubits have a gap difference $\delta \Delta / h = 12\,{\rm GHz}$ across the JJ tunnel barriers.
The goal of $\delta \Delta$ is to suppress tunneling of QPs generated by the impact; the suppression is effective if the QPs are sufficiently ``cold'' \cite{connolly_2024}, and $\delta \Delta / h$ exceeds the qubit frequency $f_{\rm q}$, see Fig.~\ref{fig:fig1}(b).

\section{Continuous coherence characterization experiment\label{sec:t2_t2e_t1}}

\begin{figure}[t]
  \begin{center}
    \includegraphics[scale = 0.97]{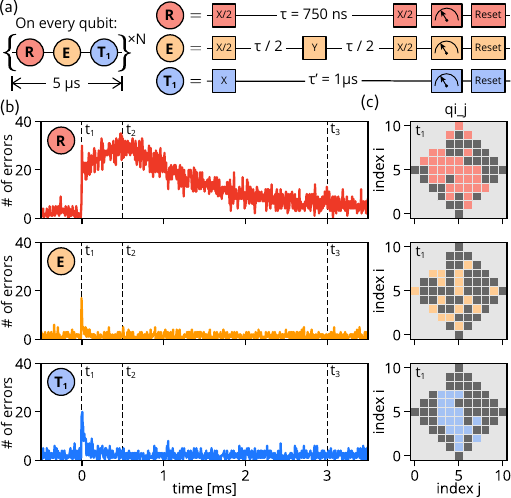}
    \caption{\textbf{Fast time-resolved sampling of qubit coherence during an impact in a gap-engineered qubit array.} \textbf{(a)} The coherence characterization sequence consists of: (i) a Ramsey measurement \textbf{R}, (ii) a spin-echo measurement \textbf{E}, and (iii) a qubit relaxation measurement \textbf{T}\bm{$_1$}.  Measurements are concatenated into a sequence, \textbf{R} + \textbf{E} + \textbf{T}\bm{$_1$}, that is repeated every 5 $\mu\text{s}$  {[see Appendix~\ref{appendix:device} for timing breakdown]}. To probe for spatial correlations, we measure 60 qubits simultaneously in a 2D array. \textbf{(b)} The total number of errors across the array for \textbf{R}, \textbf{E}, and \textbf{T}\bm{$_1$} measurements versus time during an impact event. The dynamics are separated into three timescales, {corresponding to the start of the impact ($t_1$), \textbf{E} and \textbf{T}\bm{$_1$} recovery ($t_2$), and \textbf{R} recovery ($t_3$).} \textbf{(c)}~Real-space distribution of errors at $t = t_1$.}
    \label{fig:fig2}
  \end{center} 
\end{figure}

To investigate correlated error bursts in the presence of gap engineering, we developed a measurement protocol to continuously monitor qubit coherence across the array.
The protocol has two key advancements over the previous works \cite{mcewen_2022, mcewen_2024}.
First, in addition to $T_1$ errors, we also track the phase errors.
Second, {the} spacing between measurements is {now} comparable {to} a typical QEC cycle duration, $\sim 1\,{\rm \mu s}$, whereas before it was $100\,{\rm \mu s}$~\cite{mcewen_2024}.

Specifically, on each qubit, we repeatedly perform a sequence of three measurements  {illustrated in Fig.~\ref{fig:fig2}(a)}: a Ramsey measurement \textbf{R}, a spin-echo measurement \textbf{E}, and an energy relaxation measurement \textbf{T}\bm{$_1$}.
In \textbf{R}, we initialize each qubit in a superposition state, $\ket{\psi} = (|0\rangle + i |1\rangle) / \sqrt{2}$, {free evolve} for $750\,{\rm ns}$, and apply a $\pi / 2$-pulse {followed by measurement}.
The pulse sequence in \textbf{E} is similar, except for {a refocusing} $\pi$-pulse in the middle of the {free evolution} period.
Lastly, in \textbf{T}\bm{$_1$}, we prepare each qubit in $|1\rangle$, wait for $1\,{\rm \mu s}$, and measure.
In all three measurements, {the} outcome is $1$ in the absence of dephasing and energy relaxation. 
An outcome of $0$ indicates a phase error (for \textbf{R} and \textbf{E}) or a relaxation error (for \textbf{T}\bm{$_1$}).
After each {measurement} operation, we reset the qubit to the $|0\rangle$ state. 
The {entire \textbf{R} + \textbf{E} + \textbf{T}\bm{$_1$}} sequence takes $\Delta t = {5}\,{\rm \mu s}$ (see Appendix~\ref{appendix:device} for a timing breakdown).
We repeat it $N = 8 \cdot 10^5$ times to form a single dataset that continuously tracks the {dynamics of qubit} coherence over a period of $N \Delta t = {4}\,{\rm sec}$.
We collect a total of 1800 such datasets (${2}\,{\rm hours}$ of data acquisition).

To locate error bursts in the data, we compute the total number of errors across the array in each measurement cycle, and record how this number changes in time. 
{We isolate bursts from the background noise by applying a matched filter to the error number time trace.}
Details on the matched-filter template and detection procedure are provided in Appendix~\ref{appendix:analysis}. 
In total, we observe 101 error bursts corresponding to an average burst rate of $1 / (71\,\text{sec})$ \footnote{We note that only the largest of these bursts will give rise to errors in a high distance repetition code. This is why logical errors in Ref.~\cite{google_2024} have a smaller rate.}.

Figure \ref{fig:fig2}(b) shows a representative error burst (see Fig.~S1 in Ref.~\cite{supplement} for additional examples).
A sudden, correlated increase in the number of errors occurs simultaneously in \textbf{R}, \textbf{E}, and \textbf{T}\bm{$_1$} measurements, and is followed by {a recovery} to the baseline level.
 {These errors are correlated not only in time but also in space, see Fig.~\ref{fig:fig2}(c). 
Ramsey errors are the most pronounced, both in terms of the number of {qubits} affected and the {recovery time}. 
This shows that} the dominant type of errors during the radiation impacts are {correlated} {\it phase errors}, persisting at an elevated level for $\sim 1\,{\rm ms}$.
Remarkably, apart from the initial {$\sim 10$}~$\mu s$ period at the beginning of a burst, the phase errors disappear under the spin-echo sequence.
This indicates that they are primarily caused by {\it quasi-static} qubit frequency shifts.
The observation of quasi-static frequency shifts during the impact events is the central finding of this work. 

In addition to the burst of phase errors, there is also a transient $T_1$ error burst. 
This burst lasts on the order of $10\,{\rm \mu s}$.
The latter timescale is smaller than the duration of $T_1$ error bursts on devices without gap engineering \cite{mcewen_2022} by more than two orders of magnitude. {For additional information about $T_1$ error bursts, see Appendix \ref{appendix:fast_T1}.}

\section{Quasiparticle-induced qubit frequency shifts\label{sec:tomo}}

\begin{figure}[t]
  \begin{center}
    \includegraphics[scale = 1]{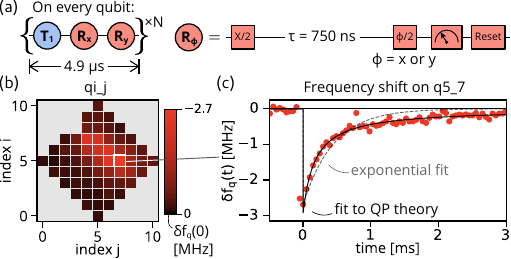}
    \caption{\textbf{Investigating QP-induced frequency shifts using time-resolved tomography.} \textbf{(a)} The control sequence consists of: (i) the \textbf{T}\bm{$_1$} measurement from Fig.~\ref{fig:fig2}(a), (ii-iii) Ramsey measurements with a final half-rotation about the $x$- and $y$-axis, respectively (\bm{$\mathrm{R}_\mathrm{X}$}, \bm{$\mathrm{R}_\mathrm{Y}$}). The measurements are concatenated into a sequence, \textbf{T}\bm{$_1$} + \bm{$\mathrm{R}_\mathrm{X}$} + \bm{$\mathrm{R}_\mathrm{Y}$}, that is repeated every 4.9 $\mu\text{s}$. We estimate the instantaneous frequency deviation of each qubit{, $\delta f_{\rm q}(t)$, using the average of ten subsequent \bm{$\mathrm{R}_\mathrm{X}$} \& \bm{$\mathrm{R}_\mathrm{Y}$} measurements.} \textbf{(b)} A heat map of the peak value of $\delta f_{\rm q}(0)$ across the 60 qubit grid during an impact.
    \textbf{(c)}~Dynamics of $\delta f_{\rm q}(t)$ for a single qubit{, q5\_7,} during the impact of panel (b). The time-evolution of {$\delta f_{\rm q}$} is consistent with a theory in which $\delta f_{\rm q} (t) / f_{\rm q} \propto x_{\rm qp}(t)$, where $x_{\rm qp}(t)$ is governed by recombination processes [see Eqs.~\eqref{eq:shift} and \eqref{eq:kinetics}].
   }
    \label{fig:fig3}
  \end{center}
\end{figure}

In order to capture magnitude, sign, and dynamics of the {quasi-static} frequency shifts, we perform a time-resolved tomography experiment. 
We use this experiment to verify the QP-based origin of the error bursts.

\subsection{Tomography experiment}

We implement the tomography by repeating a sequence of three measurements on each qubit: \textbf{T}\bm{$_1$}, \bm{$\mathrm{R}_\mathrm{X}$}, and \bm{$\mathrm{R}_\mathrm{Y}$}~\footnote{We include the \textbf{T}\bm{$_1$} measurement to collect additional statistics on the qubit relaxation errors; this measurement is not used in the frequency shift analysis.}.
The core of the sequence are measurements \bm{$\mathrm{R}_\mathrm{X}$} and \bm{$\mathrm{R}_\mathrm{Y}$}.
These are the two variants of the Ramsey measurements that differ by the rotation axis {of the final} $\pi/2$ pulse [see Fig.~\ref{fig:fig3}(a)].
It is the difference in the axis that allows us to extract the frequency shift.
To understand how, note that both \bm{$\mathrm{R}_\mathrm{X}$} and \bm{$\mathrm{R}_\mathrm{Y}$} begin by initializing {the qubit on the equator of the Bloch sphere, $\ket{\psi} = (|0\rangle + i |1\rangle) / \sqrt{2}$}.
If the qubit frequency is shifted by $\delta f_{\rm q}$, then the qubit starts precessing around the $z$-axis in the original reference frame.
Over the free evolution {time,} $\tau$, it rotates by an angle $\varphi = 2\pi \delta f_{\rm q}\tau$.
The measurement outcomes ${\cal M}(\mathbf{R}_\mathbf{X})$ and ${\cal M}(\mathbf{R}_\mathbf{Y})$ 
provide information about $\varphi$.
Specifically, their expectation values satisfy $\langle {\cal M}(\mathbf{R}_\mathbf{X}) \rangle = {(1 - \cos \varphi) / 2}$ and $\langle {\cal M}(\mathbf{R}_\mathbf{Y}) \rangle = {(1 - \sin \varphi) / 2}$.
We {estimate} $\delta f_{\rm q} = \varphi / (2\pi \tau)$ by averaging ten subsequent measurement outcomes to extract $\varphi$.
This procedure is justified by the fact that $\delta f_{\rm q}$ changes on timescales $\sim 1\,{\rm ms}$ vastly exceeding the averaging window $\approx 50\,{\rm \mu s}$.

An example of the dependence of $\delta f_{\rm q}$ on qubit location and time during the error burst is shown in Fig.~\ref{fig:fig3}(b,c)
(see Fig.~S2 in Ref.~\cite{supplement} for additional examples and Appendix~\ref{appendix:stats} for statistical information).
There are several notable features in these data.
The first is the magnitude and sign of the frequency shifts.
The frequency shifts are in the MHz range, and are non-uniform in space.
They are the largest at the epicenter of the burst, where they reach $2.7\,{\rm MHz}$ in this example. 
Strikingly, the shifts are {\it negative} on all qubits.
{This behavior applies to all detected error bursts.} 
{Another feature that warrants attention is that the frequency recovery---which takes approximately $1\,{\rm ms}$---is poorly described by an exponential time-dependence (dashed line in Fig.~\ref{fig:fig3}(c)).} 
As we will now explain, both the non-exponential character of the recovery and the negative sign of $\delta f_{\rm q}$ are consistent with frequency shifts originating from the elevated QP density in the JJ region.

\subsection{Interpretation of ${\delta f_{\rm q}(t)}$ in terms of QPs}

Let us first outline the theory predictions for the QP influence on the qubit \cite{catelani_relaxation_2011, mcewen_2024}.
The primary mechanism responsible for the variation of the qubit frequency with the QP density is the change in the Josephson energy, $E_J$, of the qubit's JJs.
The presence of QPs near the JJ impedes the supercurrent flow.
This reduces $E_J$ and thus {\it decreases} the qubit frequency.
There are other, weaker effects that further decrease $f_{\rm q}$, such as renormalization of the junction capacitance due to the QP response.
In total, we find (see Appendix~\ref{appendix:shifts} for details):
\begin{equation}\label{eq:shift}
    \frac{\delta f_{\rm q}}{f_{\rm q}} = - a \, x_{\rm qp},
\end{equation}
where $x_{\rm qp}$ is the QP density normalized by the density of Cooper pairs.
The numeric coefficient $a$ depends on the gap difference $\delta \Delta$ and $hf_{\rm q}$; we  {theoretically} estimate $a \simeq 0.77$ for our parameters. 

Because $\delta f_{\rm q} \propto x_{\rm qp}$, the time-dependence of the shifts following the burst is controlled by dynamics of the QP density. 
At high densities, the main process responsible for the decay of $x_{\rm qp}$ near the JJ is the QP {\it recombination}. The recombination can be modelled by
\begin{equation}\label{eq:kinetics}
    \frac{d x_{\rm qp}(t)}{dt} = -r x_{\rm qp}^2(t),
\end{equation}
where $r$ is the recombination rate.
Solving this equation for $x_{\rm qp}(t)$ and combining the solution with Eq.~\eqref{eq:shift}, we find a non-exponential recovery profile,
\begin{equation}\label{eq:fq_vs_t}
    \delta f_{\rm q}(t) = \frac{\delta f_{\rm q}(0)}{1 + t / t_{\rm rec}}, \quad\quad t_{\rm rec} = \frac{1}{r}\frac{a f_{\rm q}}{\delta f_{\rm q}(0)}.
\end{equation}
Here, $\delta f_{\rm q}(0)$ is the shift value at the start of the burst.

To compare the data with the theory, we fit the measured $\delta f_{\rm q}(t)$ to the functional form in Eq.~\eqref{eq:fq_vs_t}~\footnote{In order to improve the fit precision, we allow $\delta f_{\rm q}(0)$ to vary in a narrow interval around the measured peak value of $\delta f_{\rm q}(t)$.}.
The model accurately reproduces the observed time-dependence, see Fig.~\ref{fig:fig3}(c).
From the data of Fig.~\ref{fig:fig3}(c), we extract the recombination rate of $r =  {1/(105\,\text{ns})}$. 
Combining the fit results from different qubits and different bursts, we obtain
${r=1/(88 \pm 12\,\text{ns})}$.
This value agrees with earlier works on QP kinetics in aluminum qubits~\cite{wang_2014, mcewen_2022, diamond_2022, mcewen_2024}.

We also use Eq.~\eqref{eq:shift} to obtain the QP density immediately after the impact.
For the burst shown in Fig.~\ref{fig:fig3}(b), we find $x_{\rm qp} (0) \simeq 3.2\cdot 10^{-4}$ at the epicenter. 
This number is close to the result of a recent radiation impact simulation~\cite{yelton_2024}, notwithstanding the differences in the devices considered.
All in all, the frequency shifts are consistent with expectations based on the QP theory.

\begin{figure}[t]
  \begin{center}
    \includegraphics[scale = 1]{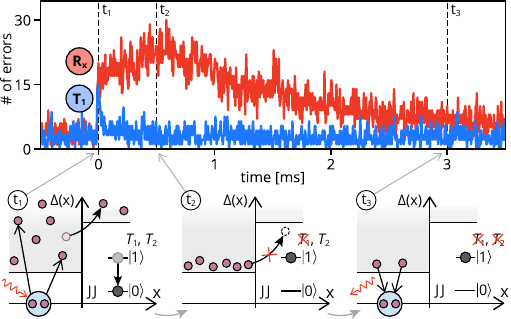}
    \caption{{\bf Interpretation of timescale separation in terms of QP kinetics.} Immediately after the impact, $t = t_1$, the QPs are energetic enough to overcome the gap difference resulting in both $T_1$- and $T_2$-errors. By $t = t_2$, the QPs have cooled down to the vicinity of the gap edge; this eliminates the $T_1$-errors {but} the $T_2$-errors {remain due to the} elevated levels of $x_\text{qp}$ near the JJs. For $t \geq t_3$, QP recombination process have led to a sufficiently dilute level of $x_\text{qp}$,  {suppressing} $T_2$-errors.}
    \label{fig:fig4}
  \end{center}
\end{figure}

\subsection{Timescale separation between phase and $T_1$ error bursts}

In addition to explaining frequency shifts, the QP kinetics offers an interpretation of the timescale separation between $T_1$ and phase errors, see~Fig.~\ref{fig:fig4}.
During an impact event, QPs are generated via the absorption of highly-energetic phonons.
Therefore, immediately after the impact, they have a broad energy distribution. 
$T_1$ errors occur as long as the characteristic QP energy exceeds $\delta \Delta - h f_{\rm q}$ \cite{diamond_2022, connolly_2024}.
The relaxation of QP energy is governed by phonon emission processes \cite{glazman2021}.
We estimate that it takes $\sim {10}\,{\rm \mu s}$ for QPs to cool down below $\delta \Delta - h f_{\rm q}$, in agreement with the observed duration of the $T_1$ error bursts, {see Appendices \ref{appendix:fast_T1} and \ref{appendix:qp_relaxation} for details}.
Unlike $T_1$ errors, the frequency shifts persist even after the QP tunneling has ceased; they disappear only when there are no QPs left near the junction.
This is why the recovery of the qubit frequencies has a much longer timescale.

The fact that the initial QP energy distribution is broad is further verified by the observation of correlated {\it excitation} errors,  {i.e., errors in which the qubit transitions from $|0\rangle$ to $|1\rangle$,} see Appendices~\ref{appendix:excitation} and \ref{appendix:qp_relaxation}.
As expected from the QP kinetics, the duration of the excitation error bursts, $\lesssim 5\,{\rm \mu s}$, is shorter than the duration of the $T_1$ error bursts.

\section{Radiation impact effects on Quantum Error Correction\label{sec:qec}}

In Secs.~\ref{sec:t2_t2e_t1} and \ref{sec:tomo}, we showed that radiation impacts cause long-lasting ($\sim 1\,{\rm ms}$) phase error bursts and transient ($\sim 10\,{\rm \mu s}$) $T_1$ error bursts.
A key remaining question is whether one or either of the two burst types affect QEC performance.

\subsection{Interleaved monitoring experiment\label{sec:interleaved}}
{To address the effect of impacts on QEC}, we perform an interleaved experiment in which we run a repetition code on one part of the qubit array, while monitoring for phase and $T_1$ errors on an adjacent part.
We correlate bursts in {QEC error detections} with the two types of error bursts on monitor qubits.

\begin{figure}[t]
  \begin{center}
    \includegraphics[scale = 1]{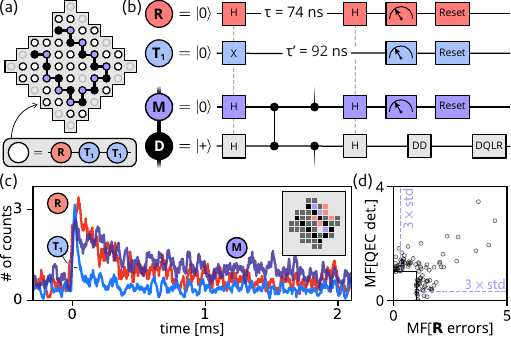}
    \caption{{\bf Interleaved repetition code and coherence characterization experiment.} {\bf (a)} Black and purple circles represent data and measure QEC qubits, respectively; white circles show the coherence monitor qubits. {\bf (b)} The repetition code and coherence characterization circuits. We use the phase flip version of the repetition code, with data qubits initialized in $|+\rangle = [|0\rangle + |1\rangle] / \sqrt{2}$. The two qubit gates used in the code are CZ gates. Coherence characterization sequence includes \textbf{R} and $\mathbf{T_1}$ measurements. {\bf (c)} Example of an error burst in \textbf{R} measurements (red), $\mathbf{T_1}$ measurements (blue), and repetition code detections (purple). {A detection refers to an event in which two subsequent {\bf M} measurements return different outcomes.} Each curve shows the total number of detections/errors across the array; to reduce the amount of noise in the data, we averaged the traces over a window of $\approx 50\,{\rm \mu s}$. The detection burst aligns closely with the error burst in \textbf{R} measurements.
    Inset: spatial distribution of QEC detections and \textbf{R} errors at the start of the burst.
    {\bf (d)} Correlation between repetition code detection bursts and error bursts in \textbf{R}. The plot shows the distribution of the values of the matched filter (MF) applied separately to QEC and \textbf{R} data, at the time of the impact. Purple dashed lines correspond to $3\sigma_{\rm MF}$, where $\sigma_{\rm MF}$ is the standard deviation of the filtered signal outside of the burst. Black solid lines represent thresholds used in the burst detection.}
    \label{fig:fig5}
  \end{center}
\end{figure}

The spatial layout and circuit diagrams for the interleaved experiment are shown in Fig.~\ref{fig:fig5}(a,b). 
We run the repetition code on the chain of qubits depicted by black and purple circles.
The repetition code is a simple QEC protocol that can only correct for a single type of error \cite{Kelly2015,chen_2021}.
Throughout the section, we focus on the X-basis code, i.e., the version of the code aimed {at correcting for phase errors} [see Appendix~\ref{appendix:qec} for the Z-basis code results]. 
The data qubits (black) store the logical information. 
In each QEC cycle, we apply a sequence of operations that is calibrated to map the joint X parity of neighboring data qubits onto the state of the measure qubit (purple);  {this achieves the measurement of the code stabilizers}.
The interval between the two subsequent Hadamard pulses in the stabilizer measurement is $74\,{\rm ns}$, same as in Ref.~\cite{google_2024}.


On monitor qubits, we repeatedly apply a sequence of three measurements: a Ramsey measurement \textbf{R} and two energy relaxation measurements $\mathbf{T_1}$. 
We synchronize each coherence measurement with the repetition code cycle in such a way that they have the same total duration ($944\,{\rm ns}$). The free evolution time of \textbf{R}, $\tau = 74\,{\rm ns}$, matches the interval between two Hadamard gates in the repetition code \footnote{An inconsequential difference in the present experiment from the experiment of Sec.~\ref{sec:t2_t2e_t1} is that we use Hadamard gates instead of $\text{X}/2$-gates in \textbf{R}.}. The wait time of $\mathbf{T_1}$ is $92\,{\rm ns}$.

We find impact events in the following way.
For monitor qubits, we record the total number of \textbf{R} errors in each cycle, and apply the matched filter to the resulting time trace.
For the repetition code, we apply the matched filter to the total number of {\it detection events} per cycle.
{A detection event occurs when two subsequent QEC measurements return opposite outcomes \cite{Kelly2015}.}
We identify a radiation impact when at least one of the two matched filter signals shows a prominent peak.
Details of the analysis---such as the choices of the matched filter thresholds---are discussed in Appendix~\ref{appendix:qec_details}. 

Let us first consider an example of a relatively strong radiation impact, and compare how the impact manifests in repetition code and monitor qubit measurements, see Fig.~\ref{fig:fig5}(c). 
Same as in Sec.~\ref{sec:t2_t2e_t1}, on monitor qubits, we observe error bursts in both \textbf{R} and $\mathbf{T_1}$ measurements.
The two bursts have different timescales: the burst in $\mathbf{T_1}$ ceases over $\sim 10\,{\rm \mu s}$, while \textbf{R} errors linger for $\lesssim 1\,{\rm ms}$.
A burst also occurs in the repetition code detections.
A striking feature of Fig.~\ref{fig:fig5}(c) is that the shape and the duration of the repetition code burst align closely with those of the \textbf{R} error burst---and not with those of the $\mathbf{T_1}$ burst.
This provides an indication that it is the phase errors that underlie the QEC detection bursts.

We collected a total of 8 hours of measurement data and found 105 bursts (see Fig.~S3 in Ref.~\cite{supplement} for additional examples of error burst traces). 
To further connect QEC bursts with phase error bursts, in Fig.~\ref{fig:fig5}(d), we show the distribution of the matched filter values in QEC detections and \textbf{R} errors at the time of the impact.
These values can be taken as proxies for the respective burst sizes.
We observe that bursts in \textbf{R} measurements are generally accompanied by repetition code bursts, with statistical significance exceeding three standard deviations.
{Similarly,} repetition code bursts are accompanied by statistically significant bursts in \textbf{R}.
We verified that borderline cases, where one of the two bursts is much larger than the other, occur when a weak impact happens in a part the array containing only one type of qubits---either monitor or QEC. 

\begin{figure}[t]
  \begin{center}
    \includegraphics[scale = 1]{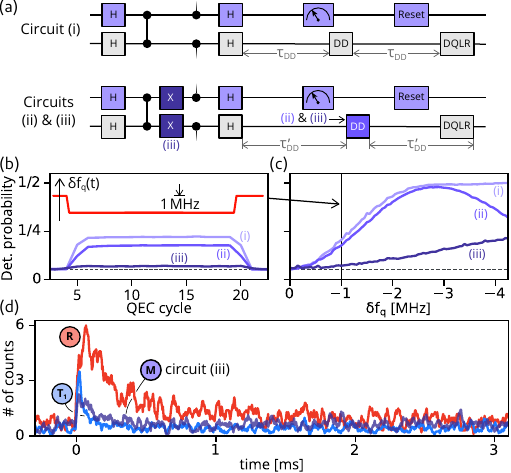}
    \caption{\textbf{Mitigation of the effect of frequency shifts on repetition code.} {\bf (a)}  We modify the standard repetition code circuit (i) in two ways to decrease the code's sensitivity to coherent phase accumulation. First, in (ii), we displace the center of the dynamical decoupling sequence to account for the duration of DQLR operation (data qubit leakage removal). Then, in (iii),  {we add an echo pulse halfway between the two Hadamards. Because phase errors commute through CZ gates, echo should efficiently cancel phase accumulation in this part of the circuit.} {\bf (b)} To quantify the sensitivity of the modified circuits to frequency shifts, we controllably inject a step-function shift $\delta f_{\rm q}(t)$ on all qubits, and measure the response of the detection probability [$\delta f_{\rm q} = -1\,{\rm MHz}$ in the plot]. 
    With each progressive modification, the repetition code responds less to the frequency shift. {\bf (c)} Dependence of the average detection probability on $\delta f_{\rm q}$. 
    {\bf (d)} Example of repetition code detections in circuit (iii) after the radiation impact (purple). Red and blue curves depict errors in \textbf{R} and $\mathbf{T_1}$ measurements on the monitor qubits [see Fig.~\ref{fig:fig5}(a) for the layout]. The duration and shape of the detection burst in circuit (iii) aligns closely with errors in $\mathbf{T_1}$ measurements.}
    \label{fig:fig6}
  \end{center}
\end{figure}

\subsection{Frequency shift injection experiment\label{sec:injection}}

{As we showed in Secs.~\ref{sec:t2_t2e_t1} and \ref{sec:tomo}, phase error bursts originate from impact-induced frequency shifts.}
To characterize the sensitivity of our QEC circuit to frequency shifts in a controlled, quantitative way---and explore methods of reducing this sensitivity---we perform an additional experiment. 
We utilize flux-tunability of our qubits and {\it deterministically} inject frequency shifts into the repetition code.
We study how a spatially uniform shift $\delta f_{\rm q}$ affects the mean probability of repetition code detections.

First, we focus on the original repetition code circuit, see circuit (i) in Fig.~\ref{fig:fig6}(a).
We apply a step-function $\delta f_{\rm q}(t)$ with amplitude $-1\,{\rm MHz}$ for the duration of $15$ QEC cycles. 
We observe that the detection probability rises by $17\%$ for approximately the same duration, see curve (i) in Fig.~\ref{fig:fig6}(b).
We also vary the step-function amplitude and track the detection probability response [curve (i) in Fig.~\ref{fig:fig6}(c)]. The detection probability rises monotonically in a $4\,{\rm MHz}$ interval of frequency shifts, approaching $50\%$ at the upper end of the interval.
This verifies that the frequency shifts of the order of those measured in tomography experiments [see Sec.~\ref{sec:tomo} and Appendix~\ref{appendix:stats}] pose danger for QEC.

The sensitivity of QEC to quasi-static frequency shifts can be reduced by improving dynamical decoupling in the circuit.
We identify two parts of the circuit that are susceptible to phase accumulation errors.
The first is the data qubit leakage removal operation (DQLR).
Although circuit (i) has a dynamical decoupling sequence on data qubits, this sequence only accounts for the durations of measure and reset operations, leaving DQLR unaccounted for.
The second part is between the two Hadamard gates on measure qubits.
In this region, the measure qubits are on the equator of the Bloch sphere.

To solve the first problem, in circuit (ii), we recenter the dynamical decoupling sequence to account for DQLR.
To solve the second problem, in circuit (iii), we include an echo pulse between Hadamards. 
Circuits (ii) and (iii) are depicted in Fig.~\ref{fig:fig6}(a).

The dependence of the detection probability on $\delta f_{\rm q}$ for the modified circuits is shown in Fig.~\ref{fig:fig6}(c).
We observe that each subsequent modification reduces the susceptibility of the repetition code to frequency shifts.
An interesting feature of the results for circuit (ii) is that the detection probability {decreases} for $|\delta f_{\rm q}| > 2.9\,{\rm MHz}$.
This downturn reflects the coherent nature of phase errors between Hadamards, which is the only part of circuit where appreciable phase accumulation remains.
Addition of echo pulses in circuit (iii) counteracts this phase accumulation resulting in a further, significant suppression of the detection probability.
In particular, the detection probability is only $2\%$ above the background at $\delta f_{\rm q} = -1\,{\rm MHz}$ [compare with $17\%$ for circuit (i)]. We attribute the remaining degradation of QEC to gate errors induced by  frequency shifts, see Appendix~\ref{appendix:inj_modeling}.

\subsection{QEC detection bursts in a modified circuit}
Modification of the QEC circuit resulted in a strong suppression of the QEC susceptibility to the ``artificially'' injected frequency shifts. 
This provides a motivation to check how circuit (iii) handles the radiation impacts. 

An example of a QEC detection trace during the impact is shown for circuit (iii) in Fig.~\ref{fig:fig6}(d), alongside with \textbf{R} and $\mathbf{T_1}$ measurements on monitor qubits.
Remarkably, the burst in repetition code detections is now much shorter than the \textbf{R} error burst.
This in a stark contrast with circuit (i), where Ramsey and detection bursts had similar timescale [cf.~Fig.~\ref{fig:fig5}(c) and Fig.~\ref{fig:fig6}(d)].
In fact, for circuit (iii), the detection burst aligns more closely with the error burst in $\mathbf{T_1}$ measurements.
This is an expected behavior:
improvements in the dynamical decoupling reduced the circuit's sensitivity to the impact-induced frequency shifts.
The main culprit of the QEC detections in circuit (iii) are thus the qubit energy relaxation errors.
We provide additional examples of error bursts for circuit (iii) in Ref.~\cite{supplement}, see Fig.~S4.

\section{Conclusions}

Using an array of gap-engineered transmon qubits, we have uncovered a new type of correlated error caused by ionizing radiation impacts, namely, correlated phase errors [see Fig.~\ref{fig:fig2}]. 
By performing time-resolved tomography experiments, we attributed the phase errors to quasi-static shifts of the qubit frequencies persisting for $\sim 1\,{\rm ms}$ after the impacts [see Fig.~\ref{fig:fig3}].
We found a good agreement between the observed frequency shift dynamics and the theory of QP-qubit interactions.
Prior to gap engineering, the frequency shifts were masked by the suppression of qubit $T_1$, making them difficult to detect.

During a typical impact event, the peak frequency shift is $|\delta f_{\rm q}| \approx 2\,{\rm MHz}$, see Fig.~\ref{fig:fig3} and Appendix~\ref{appendix:stats}. 
The phase errors that stem from shifts of this magnitude can be detrimental for QEC.
To demonstrate this, we used the frequency tunability of our qubits and deterministically injected frequency shifts into the repetition code experiment.
We found that, for the QEC circuits similar to those in Ref.~\cite{google_2024}, a spatially uniform $2\,{\rm MHz}$ shift results in a detection probability increase by~$\approx 35\%$. 

To directly correlate QEC detections and phase errors during the impacts, we ran a repetition code experiment on a part of the qubit array, while simultaneously monitoring for qubit errors on an adjacent part. 
We observed that bursts in the QEC detections and phase error bursts happen concurrently, and have similar durations, see Fig.~\ref{fig:fig5}.
Therefore, correlated phase error bursts provide a plausible explanation for the origin of the repetition code LER floor observed in Ref.~\cite{google_2024}. 

Because frequency shifts change slowly in time, the phase errors they induce can be suppressed by improving dynamical decoupling used in the QEC circuit.
We developed a modified repetition code circuit that has a lower susceptibility to frequency shifts.
For this circuit, the detection probability raises by $5\%$ in response to a controlled injection of a $2\,{\rm MHz}$ frequency shift, see Fig.~\ref{fig:fig6}.
We find that, after the modification, the radiation impacts cause much less catastrophic error bursts, with duration of $\sim 10\,{\rm \mu s}$ only. 
We attribute the remaining errors to a different mechanism: the residual impact-induced $T_1$ degradation, see Figs.~\ref{fig:fig4} and \ref{fig:fig6}.
{We note that---although effective for the repetition code---our dynamical decoupling strategy may not apply directly to other types of QEC codes \cite{debroy2024, leroux2024, eickbusch2024}.}

{Our findings show that further hardware mitigation strategies of radiation impacts will be crucial for realization of a scalable superconducting quantum processor. For example, frequency shifts can be suppressed by implementing quasiparticle traps that would reduce the QP density near the Josephson junctions \cite{riwar_2016, riwar_2019}.
The spatial extent of the error bursts can be confined using phonon traps \cite{iaia_2022, yelton_2024}.}

\section*{Acknowledgements}
We thank the entire Google Quantum AI team for maintaining the hardware, software, cryogenics, and electronics infrastructure that enabled this experiment, and for producing an environment where this work is possible. We thank Joe Fowler (NIST) and his co-authors in Ref.~\cite{fowler_2024} for sharing numerical results of cosmic-ray modeling, making it possible to estimate the event rate as function of the minimum energy deposition threshold in our devices.  V.D.K.~acknowledges helpful discussions with L.I.~Glazman.

Alex Opremcak, Gabrielle Roberts, and Vladislav D.~Kurilovich~designed the experiments. Alex Opremcak~performed the measurements from Secs.~\ref{sec:t2_t2e_t1} and \ref{sec:tomo}, with input from Vladislav D.~Kurilovich. Gabrielle Roberts~performed the  measurements from Sec.~\ref{sec:qec}, with input from Matt McEwen, Alec Eickbusch, and Kevin Satzinger. 
Vladislav D.~Kurilovich, Gabrielle Roberts, and Alex Opremcak~analyzed the experimental data, with input of all authors.
Vladislav D.~Kurilovich~and Juan Atalaya~developed the theory of the quasiparticle-induced frequency shifts.
Lara Faoro, Lev B.~Ioffe, Vladislav D.~Kurilovich, and Igor L.~Aleiner~developed a theory in Appendix~\ref{appendix:qp_relaxation}.
Leigh S.~Martin~modeled the frequency shift injection experiment.
Alexander~Bilmes~and John Mark Kreikebaum~performed the characterization of the aluminum films.
Paul Klimov~developed the optimization algorithm for qubit frequency placement.
Andreas~Bengtsson~developed the readout optimization algorithm.
Kevin Miao~developed the reset gates.
Matthew Neeley~and Wojciech Mruczkiewicz~developed the software infrastructure.
Vladislav D.~Kurilovich, Gabrielle Roberts, and Alex Opremcak~wrote the manuscript, with contributions from Leigh S.~Martin, Matt McEwen, Lara Faoro, Lev B.~Ioffe, and Igor L.~Aleiner.
All authors contributed to revising the manuscript.

{\it Note} --- Recently, we became aware of an upcoming theoretical work \cite{antonenko2025} which studies the QP-induced frequency shifts in gap-engineered qubits. The results of this work agree with our Appendix~\ref{appendix:shifts}.

\appendix
%

\section{Device and experimental information\label{appendix:device}}
We use the 72-qubit Willow processor with gap-engineered transmons qubits as originally presented in Ref.~\cite{google_2024}. In our experiments, we use a subset of 60 qubits with nearest-neighbor connectivity. The qubits and couplers use conventional Al-AlO$_x$-Al Josephson junctions, and have aluminum capacitor pads. The average distance between the qubits is $1.15\,{\rm mm}$. The qubit chip is bump-bonded to a carrier chip that contains qubit control wiring, readout and reset circuitry, and signal connections to the sample package. 

Qubit operating frequencies are placed using the Snake optimization algorithm, which maximizes the fidelity of qubit operations in a 2D grid based on a large number of trade offs \cite{klimov_2020, klimov_2024}.
A crucial component of our experiments is the ability to simultaneously read out the entire 60-qubit grid while maintaining high readout fidelities. This ability is provided by the readout Snake algorithm~\cite{bengtsson_2024}.

The time resolution in our experiments hinges on the ability to readout and reset qubits in a fast, repetitive manner. Readout is based on the Purcell-filtered readout scheme developed in Ref.~\cite{jeffrey_2014}, with a total duration of 600 ns. Reset is based on the multi-level reset gate developed in Refs.~\cite{mcewen_2021, miao_2023}, with a total duration of 160 ns. Readout and reset operations are appended to the end of each measurement sequence, comprising a total time of 760 ns; the average assignment fidelity, $[P(0|0) +P(1|1)]/2$, is $\approx99$\% during simultaneous operation. Microwave qubit rotations, both $\pi$- and $\pi/2$-rotations, have a total duration of 25 ns and rely on conventional pulse shaping techniques \cite{motzoi_2009, chow_2010, chen_2016}. In every measurement sequence, the sampling rate (e.g., {$\Delta t=5\,\mu\text{s}$} in Fig.~\ref{fig:fig2}) was chosen to minimize the amount of dead time on the device, leading to a short period ($\sim100~\text{ns}$) where qubits dwell before the sequence is repeated.

\section{Analysis of error burst data\label{appendix:analysis}}
In this Appendix, we provide details on our procedure for analysis of the error burst data used in Secs.~\ref{sec:t2_t2e_t1} and \ref{sec:tomo}. 

All of the measurements, with the exception of \bm{$\mathrm{R}_\mathrm{Y}$} from Fig.~\ref{fig:fig3}, were designed to find the qubits in $\ket{1}$ when no errors occur (i.e., $\tau \ll T_1,~T_2$). Conversely, a measurement outcome of $\ket{0}$ indicates an error.
Since we are interested in studying correlated errors, in the spatiotemporal sense, we use the sum of all qubits found in $\ket{0}$ at each time sample as our figure of merit, and focus on statistically meaningful deviations above the background fluctuations to herald ionizing radiation impacts in the device. 
We use the sum of errors in \bm{$\mathrm{R}_\mathrm{X}$} measurements, $\Sigma(t)$, to identify these impacts due to its prominence and repeatable decay profile, see Fig.~\ref{fig:fig2}(b).
In passing, we note that the \textbf{R} and \bm{$\mathrm{R}_\mathrm{X}$} measurements are identical (compare Fig.~\ref{fig:fig2}(a) and Fig.~\ref{fig:fig3}(a)), with the subscript \bm{$\mathrm{X}$} introduced for shorthand use in our tomography experiments.

\begin{figure}[t]
\begin{center}
\includegraphics[scale = 1.0]{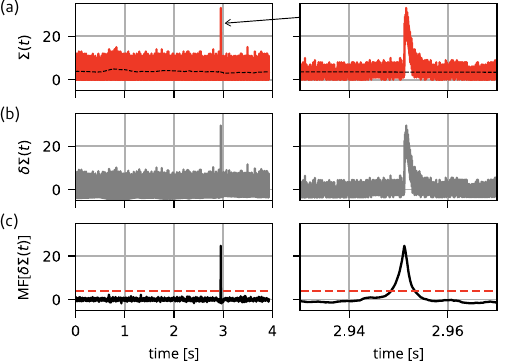}
\caption{{\bf Signal processing procedure for detecting error bursts.} 
\textbf{(a)} The sum of Ramsey errors versus time, $\Sigma(t)$, with an error burst at $t \approx 2.95~\text{s}$ (left) and a zoom-in about the burst (right). The dashed black line shows the moving average of $\Sigma(t)$ over a $\pm 25~\text{ms}$ window about each time sample.
\textbf{(b)} The deviation in $\Sigma(t)$ with respect to the moving average versus time, $\delta \Sigma(t)$ (left) and a zoom-in about the burst (right). 
Subtraction of the moving average removes components of the error signal varying on $50\,{\rm ms}$ scale.
\textbf{(c)} The matched filter applied to $\delta \Sigma(t)$ (left) with a zoom-in about the burst (right). The matched filter threshold is shown as a dashed horizontal line; when $\text{MF}[\delta \Sigma(t)]$ crosses this threshold, a potential error burst has been detected.}
\label{fig:fig_app_MF}
\end{center}
\end{figure}

A complication in impact detection comes from an appreciable level of noise in the $\Sigma(t)$ data, see Fig.~\ref{fig:fig_app_MF}(a).
The noise has a number of sources including the qubit dephasing and imperfect calibration of \bm{$\mathrm{R}_\mathrm{X} = \mathrm{R}$} sequence. 
The mean of $\Sigma(t)$ is $\langle \Sigma(t)\rangle = 2.8$ and its standard deviation is $\sigma_{\Sigma} = 1.6$.
To improve our ability to discern impacts within the noisy background, we process the $\Sigma(t)$ data via a two-step filtering procedure.
First, we compute a moving average of $\Sigma(t)$ over a $\pm 25~\text{ms}$ window about $t$ [dashed line in Fig.~\ref{fig:fig_app_MF}(a)] and subtract it from the full signal $\Sigma(t)$; we denote the result of the subtraction by $\delta\Sigma(t)$, see Fig.~\ref{fig:fig_app_MF}(b).
Then, we apply a matched filter to $\delta \Sigma(t)$.
{We use an exponential matched filter template,
\begin{equation}\label{eq:app_template}
    {\rm Template}(t) = \frac{2}{\tau_{\rm MF}}\cdot{\rm exp} (- t / \tau_{\rm MF})\,\Theta(t),
\end{equation} 
where $\Theta(t)$ is a Heaviside step function. We introduced the factor $2 / \tau_{\rm MF}$ for convenience; this factor ensures a unit normalization in the case of perfect matching, i.e., $\int_0^\infty dt\,{\rm exp}(-t / \tau_{\rm MF})\cdot {\rm Template}(t)  = 1$.
For the time constant, we pick $\tau_\text{MF} = 1.5~\text{ms}$.
Impact events are identified when the matched filter signal, $\text{MF}[\delta \Sigma(t)]$, exhibits a deviation of $\geq 4$ ``qubits'', see the dashed line in Fig.~\ref{fig:fig_app_MF}(c).}

Let us elaborate on the choice of the matched filter parameters.
Our primary motivation for using the exponential template is its simplicity. 
The actual burst profiles may deviate from an exponential shape, see Fig.~\ref{fig:fig2}(b) for an example.
However, we find that, even in these cases, the matched filter peak positions are within {$\lesssim 100\,{\rm\mu s}$} of the burst start times.

We choose the timescale $\tau_{\rm MF}$ to approximate the characteristic burst duration.
We estimate this duration by focusing on ``large'' bursts, i.e., the bursts that can be found {\it without} application of the matched filter.
For example, by looking for peaks with $\Sigma(t) \geq 20$, we find $18$ bursts in the data of Sec.~\ref{sec:t2_t2e_t1}, with durations in the interval $1-2$~ms.
Thus, we take $\tau_{\rm MF} = 1.5\,{\rm ms}$.
This choice is adequate for capturing small-size bursts (that is, bursts with size comparable to the noise level in $\Sigma(t)$) provided the durations of the latter are of the order of $1\,{\rm ms}$.
We do indeed find a large number of such small bursts.
For instance, the use of the matched filter in Sec.~\ref{sec:t2_t2e_t1} yielded $101$ bursts (compare with $18$ large bursts identified in the same data without the matched filter).
With that said, we cannot rule out that our procedure misses small bursts with timescales vastly different from $\sim 1\,{\rm ms}$.

We set ${\rm MF}[\delta \Sigma(t)]_{\rm th} = 4$ for the matched filter threshold. 
This is the lowest value of the threshold at which the identification of impact events is reliable given the amount and character of noise in the match-filtered data.
Lowering the threshold below $4$ leads to proliferation of false positives in the event detection.
For example, using ${\rm MF}[\delta \Sigma(t)]_{\rm th} = 3$, we find a rate of false positives exceeding $15\%$ (versus $\lesssim 2\%$ for ${\rm MF}[\delta \Sigma(t)]_{\rm th} = 4$).

Because of the aggressive choice of the matched filter threshold, we miss ionizing radiation impacts that deposit small amounts of energy ($\ll 100\,{\rm keV}$) into the system.
We estimate the energy resolution of our experiment in the following way.
A typical burst affects $N_{\rm q} \approx 15$ qubits shifting the qubit frequencies, see Appendix~\ref{appendix:stats}.
For simplicity, let us assume a uniform distribution of frequency shifts $\delta f_{\rm q}$. 
The expected number of simultaneous Ramsey errors due to the shifts is $N_{\rm q} \cdot [1-\cos(2\pi\delta f_{\rm q}\tau)]/2$, where $\tau = 750\,{\rm ns}$ is the free evolution time in the Ramsey sequence.
The number of errors equals four if $\delta f_{\rm q} = 230~\text{kHz}$.
By converting $\delta f_{\rm q}$ into the QP density via  Eq.~\eqref{eq:shift} and attributing an energy of $E \approx \Delta \approx 180\,{\rm \mu eV}$ to each QP, we find that the total energy in the QP system is $\approx 75\,{\rm keV}$ for an event at the detection threshold ($\Delta$ is the superconducting gap of aluminum).

\section{Statistics of bursts in tomography experiment \label{appendix:stats}}
\begin{figure}[t]
\begin{center}
\includegraphics[scale = 1]{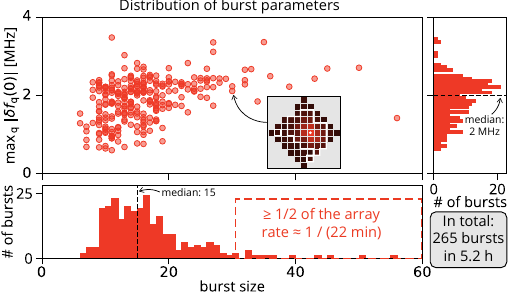}
\caption{{\bf Distribution of the burst sizes and peak frequency shifts, $\bm{{\rm max}_{\rm \,q}|\delta f_{\rm q}(0)|}$.} We define the burst size as the number of qubits with $|\delta f_{\rm q}(0)| \gtrsim {200~\text{kHz}}$ [qubits included in the burst are encircled by a white contour in the inset; white dot is a burst epicenter].
The statistics were collected over {5.2 hours} of integrated sampling time on the 60 qubit grid.}
\label{fig:fig_app_stats}
\end{center}
\end{figure}

In the tomography experiments of Sec.~\ref{sec:tomo}, we collected a total of $5.2\,{\rm h}$ of data and detected $265$ error bursts. The measured burst rate of $1 / (71\,{\rm sec})$ is smaller than the burst rate on comparable-size devices without gap engineering by a factor of $\approx 10$.
This suggests that, with gap engineering, we are only sensitive to impacts that deposit large amounts of energy, which in turn generate sufficiently high QP densities to cause measureable frequency shifts.
The measured event rate is consistent with the integrated event rate for all energy depositions {$\geq0.18$ MeV} using the simulation data from Fig. 4 of Ref.~\cite{fowler_2024}.

Figure~\ref{fig:fig_app_stats} shows a distribution of burst parameters.
The median burst size is $15$ qubits. [We define the burst size as the number of qubits with $|\delta f_{\rm q}(0)| > 3\sigma_{f} \approx 200\,{\rm kHz}$, where $\sigma_{f}$ is the RMS fluctuation of the qubit frequency outside of the burst].
However, the distribution has a long tail; there are events affecting the majority ($> 1/2$) of qubits.
Such large bursts happen with a rate of $1 / ({22}\,{\rm min})$. 
This rate is comparable to the rate of error bursts in the recent QEC experiment \cite{google_2024}.

The distribution of the peak frequency shifts during the bursts  has a median of ${2}\,{\rm MHz}$.
We expect, though, that there are many impact events with small frequency shifts ($\ll 1\,{\rm MHz}$) which we do not observe due to the insufficient measurement resolution.

\section{Correlated qubit relaxation during an error burst \label{appendix:fast_T1}}
\begin{figure}[t]
\begin{center}
\includegraphics[scale = 1]{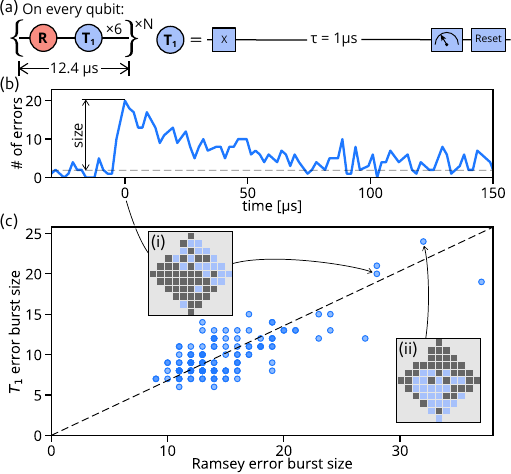}
\caption{{\bf Distribution of the $T_1$ error burst sizes and durations.} 
\textbf{(a)} Measurement sequence used in the experiment; $\mathbf{R}$ measurement is the same as in Sec.~\ref{sec:t2_t2e_t1}.
\textbf{(b)} Example of a $T_1$ error burst. 
Dashed line shows the mean background, $b = 1.9$.
\textbf{(b)} Statistics of burst sizes in \textbf{T}\bm{$_1$} and $\mathbf{R}$ measurements collected over {3.3 hours} of sampling time. 
We define the burst size as the maximum number of simultaneous errors that occurred per measurement round.
}
\label{fig:fig_app_t1}
\end{center}
\end{figure}

Here, we present the results of an additional experiment aimed at characterization of the $T_1$ error bursts.
To capture these bursts with high resolution, we designed a sequence that minimizes the spacing between subsequent energy relaxation measurements.
The sequence consists of six back-to-back \textbf{T}\bm{$_1$} measurements {preceded by} a Ramsey measurement \textbf{R}, see Fig.~\ref{fig:fig_app_t1}(a).
The average spacing between the \textbf{T}\bm{$_1$} measurements is approximately $2\,{\rm \mu s}$; it is shorter than the $5\,{\rm \mu s}$ spacing in the experiments of Secs.~\ref{sec:t2_t2e_t1} and \ref{sec:tomo}. 
The duration of the entire sequence is $12.4\,{\rm \mu s}$.
We form a single dataset by repeating the sequence $N = 4\cdot 10^5$ times on every qubit in the array ($N\Delta t = 5$ sec); we collect $2400$ such datasets.
{As} in the other experiments, we locate impact events by applying the matched filter to the Ramsey measurement data [see Appendix~\ref{appendix:analysis} for details].
The total number of events that we detect is 142.
For each {detected} event, we analyze the time-dependence the total number of $T_1$ errors that occurred per measurement round, $\Sigma_{T_1}(t)$.

Figure~\ref{fig:fig_app_t1}(b) shows an example of the $T_1$ error burst.
We characterize the burst by its {\it size} and {\it duration}.
We define the size as the peak value of $\Sigma_{T_1}(t)$. 
To find the burst duration, $t_{T_1}$, we fit $\Sigma_{T_1}(t)$ to
\begin{equation}
    c\cdot \exp(-(t - t_0) / t_{T_1})\Theta(t - t_0) + b,
\end{equation}
where $b = 1.9$ is the background value of the $T_1$ error count (held fixed) and $\Theta(t)$ is a Heaviside step function \footnote{We note that the actual shape of the burst deviates from an exponential one, cf.~Appendix~\ref{appendix:qp_relaxation_errors}. Nevertheless, we use an exponential template to extract the burst duration $t_{T_1}$ due to its simplicity.}.
In the example of Fig.~\ref{fig:fig_app_t1}(b), the size of the burst is $20$ qubits and its duration is $46\,{\rm \mu s}$. The spatial distribution of errors at the peak of the burst is shown in inset (i).

Figure~\ref{fig:fig_app_t1}(c) shows the sizes of all observed $T_1$ error bursts, alongside with the sizes of the Ramsey error bursts {detected concurrently}. 
(We define the latter size as the peak value of the Ramsey error count $\Sigma(t)$.) 
The median size of the $T_1$ error burst is $9$ qubits.
It is smaller than the median Ramsey error burst size by a factor of ${1.6}$.
The largest $T_1$ error burst that we observe has a size of $24$ qubits.
The spatial distribution of errors at the peak of this burst is shown in  inset (ii). 

Given the level of noise in the data, the fit of $\Sigma_{T_1}(t)$ produces a reliable estimate of the burst duration $t_{T_1}$ only for sufficiently large bursts (burst size $\geq 12$). Restricting to such bursts, we find $t_{T_1} = 35 \pm 15\,{\rm \mu s}$, where the spread corresponds to the burst-to-burst variation.

We note that the $T_1$ recovery time is more than two orders of magnitude faster than that on our previous generation of devices~\cite{mcewen_2022}.
The speedup of the $T_1$ recovery can be explained in terms of the QP energy relaxation processes in the following way.
With ``strong'' gap engineering, $\delta \Delta > h f_{\rm q}$, only the most energetic QPs can cause the qubit decay, $E > \delta \Delta - h f_{\rm q}$ [the energy $E$ is computed with respect to the gap edge in the low-gap film]. 
The characteristic QP energy $E$, however, rapidly relaxes following the impact [see Appendix~\ref{appendix:qp_relaxation} for details].
Thus, $T_1$ errors quickly disappear.
On the other hand, our earlier devices had $\delta \Delta < h f_{\rm q}$.
For such parameters, QPs can cause the qubit decay regardless of their energy; this led to a longer $T_1$ error burst duration $\approx 25\,{\rm ms}$.

The finding that the $T_1$ errors persist in the presence of gap engineering differs from the conclusion of Ref.~\cite{mcewen_2024}, and thus deserves a comment.
What allowed us to resolve the $T_1$ error bursts here is a significant increase in the sampling rate.  
The rate used in Ref.~\cite{mcewen_2024}, $1 / (100\,{\rm \mu s})$, was insufficient to reliably capture the $T_1$ error bursts lasting on the order of $10\,{\rm \mu s}$.
The sensitivity of our present experiment is further enhanced by the use of a much larger qubit array ($60$ qubits vs.~$6$ qubits in Ref.~\cite{mcewen_2024}).

\section{Correlated qubit excitation during an error burst\label{appendix:excitation}}
\begin{figure}[t]
\begin{center}
\includegraphics[scale = 1]{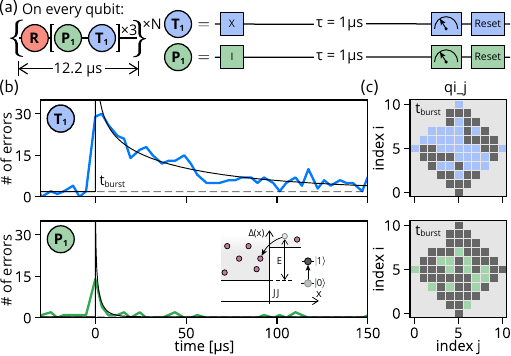}
\caption{\textbf{Correlated qubit excitation during an error burst.} 
\textbf{(a)} The excitation characterization sequence consists of: (i) a Ramsey measurement (\textbf{R}), (ii) an excitation measurement (\textbf{P}\bm{$_1$}), and (iii) a $T_1$ measurement (\textbf{T}\bm{$_1$}).
The delay time is $\tau=1~\mu\text{s}$ for both \textbf{P}\bm{$_1$} and \textbf{T}\bm{$_1$} measurements. 
At the beginning of the \textbf{P}\bm{$_1$} measurement, the qubits are prepared in $\ket{0}$.
We repeat the \textbf{P}\bm{$_1$} + \textbf{T}\bm{$_1$} measurements three times following each \textbf{R} measurement.
The \textbf{R} measurement is the same as in Sec.~\ref{sec:tomo}; we use it to identify the beginning of the burst.
The entire measurement sequence takes $12.2~\mu\text{s}$.
\textbf{(b)}~Qubit relaxation and excitation errors versus time following the radiation impact. The average number of errors in each experiment is shown as dashed horizontal line.
Black solid lines show the theory prediction [see Appendix~\ref{appendix:qp_relaxation_errors} for details]. 
Inset: Excitation errors result from tunneling of QPs with $E > \delta \Delta$ across the JJ. 
\textbf{(c)}~Real-space distributions of relaxation and excitation errors immediately after the impact, $t=t_\text{burst}$.
}
\label{fig:fig_app_t1_p1}
\end{center}
\end{figure}

{Immediately after an impact event, we expect a broad distribution of QP energies, see Fig.~\ref{fig:fig4}.}
This suggests that QP-induced qubit {\it excitation} errors {should occur}, in addition to qubit relaxation errors. 
To confirm this hypothesis, we developed a measurement sequence depicted in Fig.~\ref{fig:fig_app_t1_p1}(a).
The sequence consists of (i) a Ramsey measurement \textbf{R}, (ii) an excitation measurement \textbf{P}\bm{$_1$}, and (iii) a relaxation measurement \textbf{T}\bm{$_1$}. The \textbf{R} and \textbf{T}\bm{$_1$} measurements are the same as in Sec.~\ref{sec:t2_t2e_t1}.
The new component of the sequence is the \textbf{P}\bm{$_1$} measurement. 
In \textbf{P}\bm{$_1$}, we initialize the qubit in $|0\rangle$, wait for $1\,{\rm \mu s}$, and measure the qubit.
The outcome of $1$ means that an excitation error has occurred.
To increase the timing resolution, we repeat the \textbf{P}\bm{$_1$} + \textbf{T}\bm{$_1$} {measurement sequence} three times following each {\textbf{R}} measurement. The {Ramsey} measurement is only used to identify impact events.

As expected,  alongside Ramsey and relaxation errors, we observe a sudden increase in the number of excitation (\textbf{P}\bm{$_1$}) errors after {the} impact, see Fig.~\ref{fig:fig_app_t1_p1}(b). 
The duration of the excitation error burst is {$t_{P_1} \lesssim 5\,{\rm \mu s}$}.
It is shorter than the duration of the $T_1$ error burst by an order of magnitude.
Figure~\ref{fig:fig_app_t1_p1}(c) shows real-space distributions of relaxation and excitation errors at the peak of the burst.
The errors in \textbf{P}\bm{$_1$} and \textbf{T}\bm{$_1$} happen in the same part of the array.
However, the number of excitation errors is smaller than the number of relaxation errors by a factor of $\approx 2$.

We {attribute the} qubit excitation errors {to} QP tunneling processes {that} donate energy to the qubit.
The excitation errors occur as long as there are QPs with energy exceeding $\delta \Delta$ [here, we compute the QP energy with respect to the gap value $\Delta_L$ on the low-gap side of the Josephson junction, see {inset} in Fig.~\ref{fig:fig_app_t1_p1}{(b)}].
This interpretation allows one to explain the timescale separation between \textbf{P}\bm{$_1$} and \textbf{T}\bm{$_1$} bursts.
Indeed, the $T_1$ errors require a lower QP energy than the excitation errors, $E > \delta \Delta - h f_{\rm q}$.
It takes more time for QPs to cool down to such energies after the impact. 
We further elaborate on the theory of the QP energy relaxation in Appendix~\ref{appendix:qp_relaxation}. 
The theory predicts that the timescale separation between the error bursts in 
\textbf{P}\bm{$_1$} and \textbf{T}\bm{$_1$} is $t_{T_1} / t_{P_1} = [\delta \Delta / (\delta \Delta - h f_{\rm q})]^{9/2} \approx 20$. This value is close to the observed separation.

Due to the small number of qubits exhibiting the correlated excitation and the rapid recovery timescale for this process, we do not provide a
detailed characterization of the \textbf{P}\bm{$_1$} error bursts.
However, we note that the qualitative trends shown in Fig.~\ref{fig:fig_app_t1_p1} are representative of other bursts where correlated excitation errors were detected.


\section{Theory of excitation and relaxation error bursts \label{appendix:qp_relaxation}}
\begin{figure}[t]
\begin{center}
\includegraphics[scale = 1]{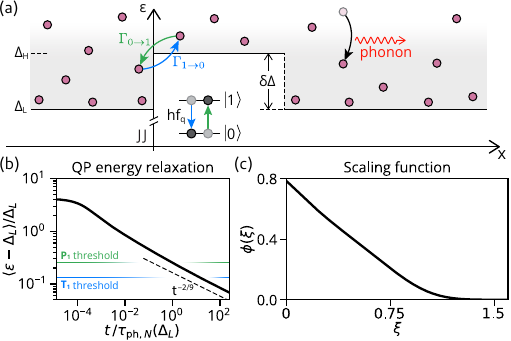}
\caption{{\textbf{Quasiparticle cooling by the phonon emission.} 
\textbf{(a)} A QP can result in the qubit excitation as long as its energy $\varepsilon$ exceeds $\Delta_L + \delta \Delta$ [$\varepsilon$ is computed  with respect to the Fermi level]. The qubit relaxation requires $\varepsilon > \Delta_L+\delta \Delta-h f_{\rm q}$. \textbf{(b)} Due to the phonon emission, the characteristic QP energy decreases in time. It approaches $\Delta_L$ as a slow power law, $\langle \varepsilon - \Delta_L\rangle / \Delta_L \propto 1 / t^{2/9}$ [$\langle \dots \rangle$ denotes averaging over the QP energy distribution $n(\varepsilon, t)$]. To produce the curve, we numerically solved Eq.~\eqref{eq:kinetic_equation} with initial condition $n(\varepsilon, t = 0) \propto \Theta(10\Delta_L - \varepsilon)$. \textbf{(c)} Scaling function determining the self-similar shape of the QP energy distribution at late times [see Eq.~\eqref{eq:scaling}].}
}
\label{fig:fig_app_t1_p1_theory}
\end{center}
\end{figure}

High-energy impacts produce ``hot'' QPs in proximity to the junction.
Hot QPs can tunnel through the junction even with gap engineering; the tunneling manifests as qubit excitation or qubit relaxation errors.
In this Appendix, we theoretically estimate the durations $t_{P_1}$ and $t_{T_1}$ of the respective error bursts, and develop a quantitative model for the qubit errors.

\subsection{Estimates of burst durations}

The burst durations $t_{P_1}$ and $t_{T_1}$ are determined by the cooling of QPs via the phonon emission.
Qubit excitation ceases when the QP energy $\varepsilon$ measured with respect to the Fermi level reduces below $\Delta_H \equiv \Delta_L + \delta \Delta$, see Fig.~\ref{fig:fig_app_t1_p1_theory}(a) [$\Delta_L$ and $\Delta_H$ denote the superconducting gap values on low- and high-gap sides of the Josephson junction, respectively].
In turn, qubit relaxation stops when $\varepsilon$ decreases to $\Delta_L + \delta \Delta - hf_{\rm q}$.
We find the time it takes for $\varepsilon$ to cross each of the two thresholds by evaluating the energy-dependence of the QP phonon emission rate.

In general, the rate for a QP can be expressed in terms of the differential phonon emission rate ${\cal F}^N_{\varepsilon-\varepsilon^\prime}$ of a normal-state electron \cite{Savich2017}:
\begin{equation}
\frac{1}{\tau_{\rm ph}(\varepsilon)}=\int_{\Delta_L}^\varepsilon \frac{\varepsilon^\prime d\varepsilon'}{\sqrt{{\varepsilon^{\prime 2}} - \Delta_L^2 }}\left[1-4u_{\varepsilon}v_{\varepsilon}u_{\varepsilon'}v_{\varepsilon'}\right]{\cal F}_{\varepsilon-\varepsilon'}^{N}. \; \label{f4}
\end{equation}
Here, the term ${\varepsilon^\prime}/{\sqrt{{\varepsilon^{\prime 2}} - \Delta_L^2 }}$ reflects the BCS singularity in the density of states, and $u_\varepsilon = \sqrt{(1 + \Delta_L/ \varepsilon) / 2}$ and $v_\varepsilon = \sqrt{(1 - \Delta_L / \varepsilon) / 2}$ are the coherence factors.
In Eq.~\eqref{f4}, we assume that the final QP states are empty.
This assumption is justified by the fact that the QP density remains relatively small during the burst, $x_{\rm qp} \ll 1$.
Throughout the Appendix, we focus on the QP cooling in the low-gap parts of the device. 
The volume of these parts is much larger than that of high-gap parts; thus, it is the former that set the QP energy distribution across the system.
This is why we use $\Delta_L$ in Eq.~\eqref{f4}.

The details of the electron-phonon interaction in a metal---and, consequently, the form of ${\cal F}_{\varepsilon-\varepsilon'}^{N}$---depends on the comparison between the electron mean free path $l$ and the phonon wavelength $\lambda_{\rm ph}$. 
To determine the electron mean free path, we measured the resistivity $\rho$ of our Al films directly above $T_{\rm c}$, $T = 1.5-2\,{\rm K}$.
Our films have $\rho=11\times10^{-9}~{\rm \Omega\,m}$ corresponding to $l=45\,\text{nm}$.
We converted $\rho$ to $l$ using the relation $\rho l = 12\pi^3 \hbar / (e^2 S_F)$, where $S_F$ is the area of the Fermi surface \cite{gall2016}.
The measured $l$ is shorter than the phonon wavelength $\lambda_{\rm ph}\sim 0.1-1\,{\rm \mu m}$ at relevant frequencies $\lesssim 10\,{\rm GHz}$.
Therefore, our films are in the dirty limit.

In this limit, electron thermalization due to transverse phonons is the dominant contribution \footnote{Contribution of longitudinal phonons to electron relaxation rate is suppressed with respect to transverse phonons contribution by a factor $(u_t / u_l)^5$ \cite{reizer1985}. Here, $u_t$ and $u_l$ are the speeds of transverse and longitudinal waves, respectively. In aluminum, $u_t / u_l \approx 0.5$.}.
The dirty limit expression for the normal-state phonon emission rate is
\begin{equation}\label{eq:rate_normal_state}
{\cal F}_{\varepsilon-\varepsilon'}^{N} = {\frac{c_{N}}{5 \pi^2}}\frac{Dp_{F}^{2}}{\rho_{m}u_{t}^{5}\hbar^5}|\varepsilon-\varepsilon'|^{3},
\end{equation}
where $p_F$ is the Fermi momentum and $D$ is the diffusion coefficient. 
Alongside the electronic properties, the rate depends on the material density $\rho_m$ and the speed of transverse waves $u_t$.
$c_N$ is a numeric factor.
Refs.~\cite{Schmid1973, reizer1985, Sergeev2000} report $c_N = 1$ while Refs.~\cite{yudson2003, Savich2017} give $c_N = 1 / 2$.
The precise value of $c_N$ is not important for an order-of-magnitude estimate of $t_{P_1}$ and $t_{T_1}$. 
 
Following the impact, hot QPs with energies $\varepsilon - \Delta_L \gtrsim \Delta_L$ emit phonons very quickly and relax to smaller energies within nanoseconds. After this initial stage all energies become $\varepsilon - \Delta_L \lesssim \Delta_L$. In this regime, the relaxation rate scales with $\varepsilon - \Delta_L$ as a power-law. Using \eqref{eq:rate_normal_state} in  Eq.~\eqref{f4}, one obtains \cite{glazman2021}
\begin{equation}\label{eq:tau_ph_approx}
    \frac{1}{\tau_{\rm ph}(\varepsilon)} = \frac{\alpha_{\rm ph}}{\tau_{{\rm ph},N}(\Delta_L)} \left(\frac{\varepsilon - \Delta_L}{\Delta_L}\right)^{9/2},
\end{equation}
where $\tau^{-1}_{{\rm ph},N}(\Delta_L) = \int_0^{\Delta_L} {\cal F}^N_{\varepsilon^\prime} d\varepsilon^\prime$ is the normal-state relaxation rate evaluated at energy $\Delta_L$, and $\alpha_{\rm ph} = 128 \sqrt{2} / 63 \approx 2.9$.

Equation~\eqref{eq:tau_ph_approx} allows us to establish a relation between durations $t_{P_1}$ and $t_{T_1}$ of excitation and relaxation error bursts that is {\it independent} of microscopic details of the system.
To see this, we recall that qubit excitation and relaxation processes have different energy thresholds. 
A QP can excite the qubit as long as its energy $\varepsilon$ exceeds $\varepsilon_{\rm th} = \Delta_L + \delta \Delta$.
The threshold is lower for the qubit relaxation processes: $\varepsilon_{\rm th} = \Delta_L + \delta \Delta - h f_{\rm q}$.
By evaluating the ratio of the phonon emission rates at the two respective thresholds, we conclude that
\begin{equation}\label{eq:t1_p1_ratio}
    \frac{t_{T_1}}{t_{P_1}} = \left [\frac{\delta \Delta}{\delta \Delta - h f_{\rm q}} \right]^{9 / 2}.
\end{equation}
For our parameters, $\delta \Delta / h = 12\,{\rm GHz}$ and $f_{\rm q} \approx 6\,{\rm GHz}$, the predicted ratio is ${t_{T_1}} / {t_{P_1}} \approx 20$. It is close to the observed ratio.

We also use Eqs.~\eqref{eq:rate_normal_state} and \eqref{eq:tau_ph_approx} to estimate $t_{P_1}$ and $t_{T_1}$ individually.
The parameters for aluminum are ${p_{F}=  1.8 \cdot 10^{-24}\,\text{kg m/s}}$, ${u_{t}=3.1\cdot10^{3}\,\text{m/s}}$, and ${\rho_{m} = 2.7\cdot10^{3}\,\text{kg/m}^{3}}$.
The estimate also depends on the electron diffusion coefficient $D$ [see Eq.~\eqref{eq:rate_normal_state}]. We compute it as $D = v_F l / 3$, where $l = 45\,{\rm nm}$ is the mean free path and $v_F = 2.0\cdot 10^6\,\text{m/s}$ is the Fermi velocity. The use of these numbers results in $t_{P_1}$ on the order of $1\,{\rm \mu s}$,
and $t_{T_1}$ on the order of $10\,{\rm \mu s}$.
These order-of-magnitude estimates agree with the data.

Further refinement of estimates can be achieved by taking into account a number of features ignored in Eq.~\eqref{eq:rate_normal_state}.
This includes deviations of the Fermi surface from a spherical one, details of the device geometry, and possibility of surface wave emission. 
Extension of theory that would include these features is beyond the scope of our work.

In our estimates, we concentrated exclusively on the dirty limit relevant to our devices. For completeness, let us note that a counterpart of Eq.~\eqref{eq:rate_normal_state} for the clean case, $\lambda_{\rm ph} \gg l$, can be found in Refs.~\cite{Sergeev2000, yudson2003}. The same references contain a discussion of a highly non-trivial crossover between clean and dirty regimes.

\subsection{Model for $\mathbf{P_1}$ and $\mathbf{T_1}$ error bursts\label{appendix:qp_relaxation_errors}}
Having obtained the estimates for $t_{P_1}$ and $t_{T_1}$, we also develop a quantitative model of qubit excitation and relaxation errors after the impact.
To this end, we solve numerically the kinetic equation for the QP energy distribution function $n(\varepsilon, t)$, and use the solution to find the probabilities of errors in $\mathbf{P_1}$ and $\mathbf{T_1}$ measurement sequences [see Fig.~\ref{fig:fig_app_t1_p1} for the definition of sequences].

Under the assumption that the distribution function is uniform in space, the kinetic equation reads
\begin{widetext}
\begin{equation}
\frac{\partial n(\varepsilon, t)}{\partial t} = 
-n(\varepsilon, t)\intop_{\Delta_L}^{\varepsilon}
{d\varepsilon'}\frac{\varepsilon^\prime}{\sqrt{\varepsilon^{\prime 2}-\Delta_L^2}} \left[1-\frac{\Delta_L^2}{\varepsilon \varepsilon'}\right]{\cal F}_{\varepsilon-\varepsilon'}^{N} + \intop_{\varepsilon}^{\infty}{d\varepsilon'}n(\varepsilon^\prime, t) \frac{\varepsilon^\prime}{\sqrt{\varepsilon^{\prime 2}-\Delta_L^2}} \left[1-\frac{\Delta_L^2}{\varepsilon \varepsilon'}\right]{\cal F}_{\varepsilon^\prime-\varepsilon}^{N}. \label{eq:kinetic_equation}
\end{equation}
The first and second terms on the right hand side represent loss and gain of QPs in a state at energy $\varepsilon$, respectively. 
In Eq.~\eqref{eq:kinetic_equation}, we only account for phonon emission. 
We also neglect the QP recombination. 
The latter approximation is justified at early times, $t \ll t_{\rm rec}$, where $t_{\rm rec}\sim 1\,{\rm ms}$ in the conditions of the burst [see the discussion around Eq.~\eqref{eq:fq_vs_t}].

Equation \eqref{eq:kinetic_equation} allows one to describe the QP cooling for any initial distribution function $n(\varepsilon, t = 0)$. 
In fact, though, details of the initial distribution matter only at very short times after the impact, $t \lesssim \tau_{{\rm ph}, N}(\Delta_L) \sim 10\,{\rm ns}$ [see the discussion after Eq.~\eqref{eq:tau_ph_approx} for the definition of $\tau_{{\rm ph}, N}(\Delta_L)$].
At later times, the characteristic QP energy becomes $\varepsilon - \Delta_L \lesssim \Delta_L$ [see Fig.~\ref{fig:fig_app_t1_p1_theory}(b)], while the distribution function acquires a self-similar shape that is {\it independent} of the initial conditions. 
To explain the origin of self-similarity, let us focus on the limit $\varepsilon-\Delta_L \ll \Delta_L$.
In this limit, the kinetic equation can be represented as
\begin{equation}\label{eq:kinetic_equation_2}
    \partial_t n(\varepsilon, t) = \frac{2\sqrt{2}}{\Delta_L^{9/2}\tau_{{\rm ph},N}(\Delta_L)}\Bigl[ -n(\varepsilon, t) \int_{\Delta_L}^\varepsilon \frac{d\varepsilon^\prime}{\sqrt{\varepsilon^\prime - \Delta_L}}(\varepsilon+\varepsilon^\prime - 2\Delta_L)(\varepsilon-\varepsilon^\prime)^3 + \int_{\varepsilon}^{\infty} \frac{d\varepsilon^\prime}{\sqrt{\varepsilon^\prime - \Delta_L}}(\varepsilon+\varepsilon^\prime - 2\Delta_L)(\varepsilon-\varepsilon^\prime)^3 n(\varepsilon^\prime, t)\Bigr].
\end{equation}
\end{widetext}
We note that the dependence of the integral kernel on $\varepsilon - \Delta_L$ and $\varepsilon^\prime - \Delta_L$ is a power law.
Because of this, Eq.~\eqref{eq:kinetic_equation_2} admits a solution in the scaling form:
\begin{equation}\label{eq:scaling}
    n(\varepsilon, t) = x_{\rm qp} \cdot \phi\left(\frac{\varepsilon - \Delta_L}{{\cal E}(t)}\right) \left(\frac{\Delta_L}{2 {\cal E}(t)}\right)^{1/2},
\end{equation}
where parameter ${\cal E}(t)$ is a power law function of time.
To find ${\cal E}(t)$, we substitute the scaling ansatz into Eq.~\eqref{eq:kinetic_equation_2} and change the variables according to $\varepsilon - \Delta_L = {\cal E}(t) \cdot \xi$ and $\varepsilon^\prime - \Delta_L = {\cal E}(t) \cdot \xi^\prime$.
After the variable change, it is easy to see that the ansatz becomes consistent if
\begin{equation}\label{eq:scale_parameter}
{\cal E}(t) = \Delta_L \left[\frac{\tau_{{\rm ph}, N}(\Delta_L)}{t}\right]^{2 / 9}.
\end{equation}
Parameter ${\cal E}(t)$ has a meaning of characteristic QP energy computed with respect to the gap edge at time $t$.

To fully specify $n(\varepsilon, t)$, one also needs to know the scaling function $\phi(\xi)$. 
We obtain $\phi(\xi)$ by numerically solving Eq.~\eqref{eq:kinetic_equation_2}.
The result is depicted in Fig.~\ref{fig:fig_app_t1_p1_theory}(c). 
The scaling function is normalized by the condition $\int_0^\infty d\xi\, \phi(\xi) / \sqrt{\xi} = 1$; this condition ensures that the total QP density for $n(\varepsilon, t)$ of Eq.~\eqref{eq:scaling} is $x_{\rm qp}$.
We verified numerically that, regardless of the initial conditions, the QP energy distribution converges to the scaling form~\eqref{eq:scaling}.

Next, we use the found $n(\varepsilon, t)$ to obtain error probabilities in $\mathbf{P_1}$ and $\mathbf{T_1}$ sequences. These probabilities can be represented as
\begin{equation}\label{eq:pt1_pp1}
    p(\mathbf{P_1}) = \Gamma_{0\rightarrow 1} \tau,\quad\quad p(\mathbf{T_1}) = \Gamma_{1\rightarrow 0} \tau,
\end{equation}
where $\Gamma_{0\rightarrow 1}$ and $\Gamma_{1\rightarrow 0}$ are qubit excitation and relaxation rates, respectively, and $\tau$ is the sequence wait time [$1\,{\rm \mu s}$ in Fig.~\ref{fig:fig_app_t1_p1}].
The QP contribution to the rates has the following form
\cite{connolly_2024}:
\begin{equation}
    \Gamma_{0 \rightarrow 1} = 4\pi f_{\rm q} S_{\rm qp}(-f_{\rm q}), \quad\Gamma_{1 \rightarrow 0} = 4\pi f_{\rm q} S_{\rm qp}(f_{\rm q}),
\end{equation}
where $f_{\rm q}$ is the qubit frequency. The information about the energy distribution of QPs is encoded in the structure factor $S_{\rm qp}(f)$:
\begin{widetext}
\begin{equation}\label{eq:S_qp}
    S_{\rm qp}(f) = \frac{2}{\Delta_L + \Delta_H}\int_{\Delta_L}^{\infty} \frac{d\varepsilon\, n(\varepsilon)[\varepsilon(\varepsilon+hf) + \Delta_L \Delta_H]}{\sqrt{\varepsilon^2 - \Delta_L^2}\sqrt{(\varepsilon + hf)^2 - \Delta_H^2}}\Theta(\varepsilon - \Delta_H + h f) + [L \rightleftarrows H],
\end{equation}
\end{widetext}
where $\Theta(x)$ is the Heaviside step function. In Eq.~\eqref{eq:S_qp}, we assumed that the distribution function is the same on the two sides of the junction.

As the QP distribution function changes in time, so do the excitation and relaxation error probabilities. 
We show the result for the time-evolution of $p(\mathbf{P_1})$ and $p(\mathbf{T_1})$ after the impact by black curves in Fig.~\ref{fig:fig_app_t1_p1}(b).
To produce these curves, we used Eqs.~\eqref{eq:scaling} and \eqref{eq:scale_parameter} in Eqs.~\eqref{eq:pt1_pp1}--\eqref{eq:S_qp}.
We also converted the probabilities into the numbers of errors across the qubit array, $\Sigma_{P_1}$ and $\Sigma_{T_1}$.
To this end, we multiplied them by a common normalization factor ${\cal N}$, $\Sigma_{P_1} = {\cal N} p(\mathbf{P_1})$ and $\Sigma_{T_1} = {\cal N} p(\mathbf{T_1})$.
We used ${\cal N}$ and $\tau_{{\rm ph}, N}(\Delta_L)$ as free parameters in comparison of the theory with the data.
We recall that $1/\tau_{{\rm ph}, N}(\Delta_L)$ has a meaning of the normal-state phonon emission rate at electron energy $\Delta_L$; it is defined after Eq.~\eqref{eq:tau_ph_approx}.
The theory accurately predicts the shape of the excitation and relaxation error bursts.
The extracted value of $\tau_{{\rm ph}, N}(\Delta_L) = 14\,{\rm ns}$ is within a factor of two of $\tau_{{\rm ph}, N}(\Delta_L) = 24\,{\rm ns}$ obtained from Eq.~\eqref{eq:rate_normal_state} by taking $c_N = 1$ \cite{Schmid1973, reizer1985, Sergeev2000} and using the material parameters listed after Eq.~\eqref{eq:t1_p1_ratio}.

\section{Theory of QP-induced frequency shifts with gap engineering\label{appendix:shifts}}
In this Appendix, we develop a theory of the QP effect on the qubit frequency.
We will justify the relation between the frequency shift and the QP density, $\delta f_{\rm q} / f_{\rm q} = - a\,x_{\rm qp}$ [Eq.~\eqref{eq:shift}], and derive an expression for the proportionality coefficient in this relation:
\begin{equation}\label{eq:a_coefficient}
    a = \frac{1}{4} + \frac{1}{4\pi}\Bigl[\sqrt{\tfrac{2\Delta}{\delta \Delta - h f_{\rm q}}} + \sqrt{\tfrac{2\Delta}{\delta \Delta + h f_{\rm q}}}\,\Bigr].
\end{equation}
Here, $\delta \Delta = \Delta_H - \Delta_L$ is the difference in superconducting gap values across the JJ, and $\Delta = (\Delta_L + \Delta_H) / 2$. 
Expression \eqref{eq:a_coefficient} is valid under the assumption that QPs are ``cold'' and predominantly reside on the low-gap side of JJs, see Fig.~\ref{fig:fig1}(b).
This assumption is justified away from the immediate start of the burst ($t \sim t_2$ in Fig.~\ref{fig:fig2}(b)).
The equation also assumes that the qubit is operated close to flux bias $\Phi = 0$.
From Eq.~\eqref{eq:a_coefficient}, we obtain $a = 0.77$ for our parameter values, $\delta \Delta / h = 12\,{\rm GHz}$, $\Delta / h = 52\,{\rm GHz}$, and $f_{\rm q} \approx 6\,{\rm GHz}$. 

Below, we present a detailed derivation of Eq.~\eqref{eq:a_coefficient}. We also extend that equation to any $\Phi$ and arbitrary QP distribution.
The theory in this Appendix generalizes the results of Ref.~\cite{catelani_relaxation_2011} to gap-engineered qubits. 

\subsection{Relation of the frequency shift to the QP admittance}

We consider a flux-tunable transmon qubit poisoned by QPs. 
The presence of QPs near JJs modifies the junctions electromagnetic response; this modification results in the shift of the qubit frequency. 
To capture the shift, we follow an approach developed in Ref.~\cite{catelani_relaxation_2011}.
First, we attribute to QPs a frequency-dependent admittance $Y_{{\rm qp}, j} (\omega)$ connected in parallel to the qubit's JJs [$j = 1, 2$ distinguishes the two junctions in the qubit].
With this load admittance, one can find the qubit frequency from the condition
\begin{equation}\label{eq:resonant_condition}
    Y_{\rm q, 0}(\omega) + \sum_{j = 1, 2}Y_{j,{\rm qp}}(\omega) = 0,
\end{equation}
where $Y_{{\rm q}, 0}(\omega) = i\omega C(1 - \omega_{\rm q}^2 / \omega^2)$ is the admittance of the transmon in the absence of QPs, and $C$ is the transmon capacitance.
To the lowest order in the QP density, we can treat the second term in Eq.~\eqref{eq:resonant_condition} as a perturbation.
This leads to the following expression for the frequency shift:
\begin{equation}\label{eq:shift_vs_ImY}
    \delta \omega_{\rm q} = -\frac{1}{2C}\sum_{j=1,2}{\rm Im}\,Y_{j, \rm qp}(\omega_{\rm q}).
\end{equation}
This result shows that we need to evaluate the reactive component of the QP admittance to find $\delta \omega_{\rm q}$.  

There are two contributions to ${\rm Im}\,Y_{j, \rm qp}(\omega)$:
\begin{equation}\label{eq:ind_and_dyn}
    {\rm Im}\,Y_{j,{\rm qp}}(\omega) = {\rm Im}\,Y^{{\rm (ind)}}_{j,{\rm qp}}(\omega) + {\rm Im}\,Y^{{\rm (dyn)}}_{j,{\rm qp}}(\omega).
\end{equation}
The first contribution reflects that the presence of QPs in the JJ region reduces the junction's Josephson coupling.
It amounts to a change in the {\it inductive} response of the junction, ${\rm Im}\,Y^{{\rm (ind)}}_{j,{\rm qp}}(\omega) \propto 1 / \omega$. 
In addition to altering the quasi-static inductive response, QPs affect the {\it dynamic} response of the JJ.
This effect is described by the second term in Eq.~\eqref{eq:ind_and_dyn} satisfying ${\rm Im}\,Y^{{\rm (dyn)}}_{j,{\rm qp}}(\omega \rightarrow 0) = 0$.
We will now focus on each of the two terms separately, and relate them to the parameters of the QP distribution.

\subsection{Inductive QP response}
Let us first address the inductive part of the response [first term in Eq.~\eqref{eq:ind_and_dyn}].
In general, the inductive response is related to supercurrent $I$ flowing through the JJ. 
For a junction biased by phase difference $\varphi$, the relation is
\begin{equation}
     Y^{{\rm (ind)}}(\omega) = \frac{2\pi}{\Phi_0}\frac{\partial_\varphi I(\varphi)}{i\omega},
\end{equation}
where $\Phi_0 = h / (2e)$ is the superconducting flux quantum [to make the notations concise, we dropped the subscript $j$; we will recover it at the end of the calculation]. In the relevant case of different superconducting gaps across the tunnel barrier, $\Delta_L < \Delta_H$, the supercurrent is given by~\cite{ambegaokar_1963}:
\begin{equation}\label{eq:I_phi}
    I(\varphi) = \frac{g_T}{e} \Delta_L \Delta_H \hspace{-0.05cm} \int_{\Delta_{L}}^{\Delta_{H}}\hspace{-0.15cm} \frac{d\varepsilon\,(1 - 2 n(\varepsilon))\sin \varphi}{\sqrt{\Delta_H^2 - \varepsilon^2}\sqrt{\varepsilon^2 - \Delta_L^2}}.
\end{equation}
Here, $g_T$ is the normal-state conductance of the junction and $n(\varepsilon)$ is the energy distribution of QPs. Energy $\varepsilon$ is computed with respect to the Fermi level.

QPs reduce the supercurrent in comparison with its value $I_0(\varphi)$ at $x_{\rm qp} = 0$, $I(\varphi) = I_0(\varphi) - \delta I_{\rm qp} (\varphi)$.
This is why they affect the inductive response. We can represent
\begin{equation}\label{eq:ind_qp}
     Y^{{\rm (ind)}}_{\rm qp}(\omega) = -\frac{2\pi}{\Phi_0}\frac{\partial_\varphi \delta I_{\rm qp}(\varphi)}{i\omega}.
\end{equation}
It follows from Eq.~\eqref{eq:I_phi} that there are two effects contributing to $\delta I_{\rm qp} (\varphi)$.
First, the presence of a QP in a state with energy $\varepsilon$ blocks the contribution of this state to the supercurrent [second term in the round brackets]. 
In addition, QPs in the JJ leads reduce the value of the superconducting gaps futher decreasing $I(\varphi)$.
We denote the two contributions by $\delta I_{\rm qp}^{(n)}$ and $\delta I_{\rm qp}^{(\Delta)}$, respectively.
The first is given by
\begin{equation}\label{eq:I_qp_n}
    \delta I_{\rm qp}^{(n)}(\varphi) = \frac{2g_T}{e} \Delta_L \Delta_H \hspace{-0.05cm} \int_{\Delta_{L}}^{\Delta_{H}}\hspace{-0.3cm} \frac{d\varepsilon\,n(\varepsilon)\,\sin \varphi}{\sqrt{\Delta_H^2 - \varepsilon^2}\sqrt{\varepsilon^2 - \Delta_L^2}}.
\end{equation}
We will analyze it assuming that the characteristic energy of QPs with respect to the gap edge is small compared to the gap difference, $E \equiv \varepsilon - \Delta_L \ll \delta \Delta$. In this case, $\delta I_{\rm qp}^{(n)}(\varphi)$ can be related to the dimensionless QP density,
\begin{equation}\label{eq:density}
    x_{{\rm qp}, L} = \frac{2}{\Delta_L}\int_{\Delta_L}^{+\infty}\frac{\varepsilon d\varepsilon\,n(\varepsilon)}{\sqrt{\varepsilon^2 - \Delta_L^2}}
\end{equation}
[i.e., the QP density normalized by the density of Cooper pairs].
Indeed, a direct comparison of integrands in Eqs.~\eqref{eq:I_qp_n} and \eqref{eq:density} at $\varepsilon - \Delta_L \ll \delta \Delta$ shows that
\begin{equation}
    \delta I_{\rm qp}^{(n)}(\varphi) = \frac{x_{{\rm qp},L}}{\pi} \sqrt{\frac{2\Delta}{\delta \Delta}}\cdot I_0(\varphi).
\end{equation}
We made a further simplifying assumption $\delta \Delta \ll \Delta = (\Delta_L + \Delta_H) / 2$ and used Eq.~\eqref{eq:I_phi} to express the result in terms of the supercurrent in the absence of QPs, $I_0(\varphi)$.

Let us now turn to the contribution $\delta I^{(\Delta)}_{\rm qp}$ associated with the reduction of the superconducting gap.
The reduction due to QPs follows from the BCS self-consistency relation, and is given by
\begin{equation}\label{eq:gap_suppression}
\Delta_\alpha = \Delta_{\alpha, 0}(1 - x_{{\rm qp}, \alpha}),
\end{equation}
where $\Delta_{\alpha, 0}$ denotes the gap value in the absence of QPs [$\alpha = L, H$].
Using Eq.~\eqref{eq:gap_suppression}, we can represent $\delta I^{(\Delta)}_{\rm qp}(\varphi)$ as
\begin{equation}
    \delta I^{(\Delta)}_{\rm qp}(\varphi) = \sum_{\alpha = L, H} x_{{\rm qp}, \alpha}\,\Delta_{\alpha, 0} \frac{\partial I(\varphi)}{\partial \Delta_\alpha},
\end{equation}
where the derivative is evaluated at $\Delta_\alpha = \Delta_{\alpha,0}$ and $n(\varepsilon) = 0$.
The use of Eq.~\eqref{eq:I_phi} in this expression together with the condition $\delta \Delta \ll \Delta_L, \Delta_H$ yields
\begin{equation}\label{eq:I_qp_Delta}
   \delta I^{(\Delta)}_{\rm qp}(\varphi) = \frac{1}{2}\Bigl(x_{{\rm qp}, L} + x_{{\rm qp}, H}\Bigr)I_0(\varphi).
\end{equation}
This expression is valid up to relative corrections linear in $\delta \Delta / \Delta \ll 1$.

Having found $\delta I_{\rm qp}^{(n)}(\varphi)$ and $\delta I_{\rm qp}^{(\Delta)}(\varphi)$ [Eqs.~\eqref{eq:I_qp_n} and \eqref{eq:I_qp_Delta}], we can obtain $Y^{{\rm (ind)}}_{\rm qp}(\omega)$ with the help of Eq.~\eqref{eq:ind_qp}. A direct substitution leads to
\begin{align}\label{eq:ImY_ind}
     Y^{{\rm (ind)}}_{\rm qp}&(\omega) = -\frac{1}{i\omega L_J} \notag\\ & \times\,\cos \varphi \Bigl(\Bigl[\tfrac{1}{2} + \tfrac{1}{\pi} \sqrt{\tfrac{2\Delta}{\delta \Delta}}\,\Bigr]x_{{\rm qp}, L} + \tfrac{1}{2}x_{{\rm qp}, H}\Bigr).
\end{align}
We expressed the result in terms of the Josephson inductance $L_J^{-1} = (2\pi / \Phi_0) \partial_\varphi I_0|_{\varphi = 0} = (2\pi / \Phi_0)^2 E_J$ unperturbed by QPs.

\subsection{Dynamic QP response}
Next, we find the dynamic part of the QPs response [second term in Eq.~\eqref{eq:ind_and_dyn}].
A convenient way to do this is to relate ${\rm Im}\,Y_{\rm qp}^{\rm (dyn)}(\omega)$ to the {\it dissipative} part of the JJ admittance \cite{catelani_relaxation_2011}.
Indeed, being a function analytic in the upper complex half-plane, the admittance $Y_{\rm qp}(\omega)$ satisfies Kramers-Kr\"onig relations. This allows one to express ${\rm Im}\,Y_{\rm qp}^{\rm (dyn)}(\omega)$ as
\begin{equation}\label{eq:ImY_dyn}
    {\rm Im}\,Y^{\rm (dyn)}_{\rm qp}(\omega) = -\frac{1}{\pi} \fint \frac{{\rm Re}\,Y_{\rm qp}(\omega^\prime) d\omega^\prime }{\omega^\prime - \omega},
\end{equation}
where $\fint$ denotes the principal value of the integral.
\begin{widetext}
\noindent The dissipative component of the admittance of a gap-engineered JJ can be obtained from the results of Refs.~\cite{diamond_2022, connolly_2024} (see, e.g., Eqs.~(S5) and (S6) of Ref.~\cite{connolly_2024}). For $\omega > 0$, the expression reads
\begin{align}
    {\rm Re}\,Y_{\rm qp}(\omega) = \frac{g_T}{\hbar \omega}\int_{\Delta_L}^{+\infty}\hspace{-0.25cm}\frac{\varepsilon\,d\varepsilon}{\sqrt{\varepsilon^2 - \Delta_L^2}}&\frac{\varepsilon+\hbar\omega}{\sqrt{(\varepsilon+\hbar\omega)^2 - \Delta_H^2}}\notag\\
    &\times M_{\varepsilon, \varepsilon + \hbar\omega}\,\Theta(\varepsilon + \hbar\omega - \Delta_H)(n(\varepsilon) - n(\varepsilon + \hbar\omega)) + [L \rightleftarrows H],\label{eq:ReY}
\end{align}
where $\Theta(x)$ is a step function and $n(\varepsilon)$ is the QP distribution function; we assume that $n(\varepsilon)$ is the same in low- and high-gap parts of the device. The matrix element $M_{\varepsilon_L, \varepsilon_H}$ depends on the phase bias $\varphi$, and is given by
\begin{equation}
    M_{\varepsilon_L, \varepsilon_H} = \Bigl(1 + \frac{\Delta_L \Delta_H}{\varepsilon_L \varepsilon_H}\Bigr)\frac{1 + \cos \varphi}{2}
    + \Bigl(1 - \frac{\Delta_L \Delta_H}{\varepsilon_L \varepsilon_H}\Bigr)\frac{1- \cos \varphi}{2}.
\end{equation}
Substituting Eq.~\eqref{eq:ReY} into Eq.~\eqref{eq:ImY_dyn}, we find
\begin{align}
    {\rm Im}\,Y^{\rm (dyn)}_{{\rm qp}}(\omega) = \frac{g_T}{\pi \hbar \omega} \fint^{+\infty}_{\Delta_L}\hspace{-0.15cm}n(\varepsilon_L) \frac{\varepsilon_L\, d\varepsilon_L}{\sqrt{\varepsilon_L^2 - \Delta_L^2}} &\fint^{+\infty}_{\Delta_H} \hspace{-0.15cm}\frac{\varepsilon_H\, d\varepsilon_H }{\sqrt{\varepsilon_H^2 - \Delta_H^2}}M_{\varepsilon_L, \varepsilon_H} \notag\\
    &\times \Bigl[\frac{1}{\varepsilon_H - \varepsilon_L + \hbar\omega} + \frac{1}{\varepsilon_H - \varepsilon_L - \hbar\omega} - \frac{2}{\varepsilon_H - \varepsilon_L}\Bigr] + [L \rightleftarrows H].
\end{align}
This expression can be simplified under the assumption that the energy of QPs with respect to the gap edge is low: $\varepsilon_\alpha = \Delta_\alpha + E_\alpha$ with $E_\alpha \ll \Delta_\alpha$.
In this case, it is possible to approximate the matrix element by {$M_{\varepsilon_H, \varepsilon_L} = 1 + \cos \varphi$} and calculate one of the two energy integrals explicitly. The result of the calculation is 
\begin{align}\label{eq:ImY_dyn_general}
    {\rm Im}\,Y^{\rm (dyn)}_{{\rm qp}}(\omega) =  \frac{g_T\sqrt{\Delta_L \Delta_H}}{\hbar\omega} &\frac{1 + \cos \varphi}{2} \int^{+\infty}_{0}\hspace{-0.15cm}n(\Delta_L + E_L) \frac{dE_L}{\sqrt{E_L}} 
   \notag \\  &\times \Bigl[\frac{\Theta(\delta \Delta - \hbar\omega - E_L)}{\sqrt{\delta \Delta - \hbar\omega - E_L}} + \frac{\Theta(\delta \Delta + \hbar\omega - E_L)}{\sqrt{\delta \Delta + \hbar\omega - E_L}} - 2 \frac{\Theta(\delta \Delta - E_L)}{\sqrt{\delta \Delta - E_L}}\Bigr] + [L \rightleftarrows H].
\end{align}
To proceed, we will assume that the characteristic energy of QPs is small not only in comparison with $\Delta_{L, H}$, but also in comparison with the dissipation threshold, $E_L \ll \delta \Delta - \hbar \omega$. This allows us to neglect $E_L$ in the square brackets and express the integral in terms of the QP density [cf.~Eq.~\eqref{eq:density}]:
\begin{equation}\label{eq:ImY_dyn_final}
    {\rm Im}\,Y^{\rm (dyn)}_{{\rm qp}}(\omega) =   \frac{1}{\omega L_J} \frac{1 + \cos \varphi}{2}\,\frac{x_{{\rm qp}, L}}{2\pi}
    \Bigl[\sqrt{\tfrac{2\Delta}{\delta \Delta - \hbar\omega}} + \sqrt{\tfrac{2 \Delta}{\delta \Delta + \hbar\omega}} - 2 \sqrt{\tfrac{2 \Delta}{\delta \Delta}}\,\Bigr].
\end{equation}
Here, we used Ambegaokar-Baratoff relation to express the normal-state conductance $g_T$ in terms of the Josephson inductance, $L_J^{-1} = \pi g_T \Delta / \hbar$ \footnote{This form of the Ambegaokar-Baratoff relation is valid for $\delta \Delta \ll \Delta$ \cite{ambegaokar_1963}.}. 
We note that, under the made assumptions, ${\rm Im}\,Y^{\rm (dyn)}_{{\rm qp}}(\omega)$ depends only on the QP density on the low-gap side of the JJ, $x_{{\rm qp}, L}$.
Another interesting observation is that the low-frequency behavior of ${\rm Im}\,Y^{\rm (dyn)}_{{\rm qp}}(\omega)$ is linear in $\omega$,
\begin{equation}\label{eq:ImY_dyn_cap}
    {\rm Im}\,Y^{\rm (dyn)}_{{\rm qp}}(\omega) =  \frac{3\omega}{4L_J(\delta \Delta / \hbar)^2} \frac{1 + \cos \varphi}{2} 
    \frac{x_{{\rm qp}, L} }{2\pi}\sqrt{\frac{2 \Delta}{\delta \Delta}}\hspace{-0.25cm}.
\end{equation}
This allows one to interpret ${\rm Im}\,Y^{\rm (dyn)}_{{\rm qp}}(\omega)$ as renormalization of the junction capacitance by the QPs in the low-frequency regime, $\omega \ll \delta \Delta / \hbar$. We note, though, that our qubits have $\omega_{\rm q} \approx (\delta \Delta / \hbar) / 2$, so there are appreciable corrections to Eq.~\eqref{eq:ImY_dyn_cap}; we use the full expression, Eq.~\eqref{eq:ImY_dyn_final}, below.

Combining Eqs.~\eqref{eq:ImY_ind} and \eqref{eq:ImY_dyn_final} (and recovering index $j$ that distinguishes the two JJs in the qubit), we find for the reactive part of the QP admittance:
\begin{equation}\label{eq:ImY_final}
    {\rm Im}\,Y_{j, \rm qp}(\omega) = \frac{1}{\omega L_{Jj}} \Bigl( \tfrac{1}{2}\bigl[x_{{\rm qp}, L} + x_{{\rm qp}, H}\bigr]\cos \varphi_j + \tfrac{1 + \cos \varphi_j}{2} \tfrac{x_{{\rm qp},L}}{2\pi} \Bigl[\sqrt{\tfrac{2\Delta}{\delta \Delta - \hbar\omega}} + \sqrt{\tfrac{2 \Delta}{\delta \Delta + \hbar\omega}}\,\Bigr] - \tfrac{1 - \cos \varphi_j}{2} \tfrac{x_{{\rm qp},L}}{\pi} \sqrt{\tfrac{2   \Delta}{\delta \Delta}}\Bigr).
\end{equation}
For simplicity, here we assumed that $x_{{\rm qp}, L}$ and $x_{{\rm qp}, H}$ are the same for the two junctions of the qubit.
This expression is valid up to relative corrections that are $O(\delta \Delta / \Delta)$.
Using the found ${\rm Im}\,Y_{j, \rm qp}(\omega)$ in Eq.~\eqref{eq:shift_vs_ImY}, we arrive to a final relation between the frequency shift and the QP densities:
\begin{equation}\label{eq:freq_shift_final}
    \delta \omega_{\rm q} / \omega_{\rm q}(\Phi) = -\Bigl[\tfrac{1}{4} + \tfrac{1}{8\pi}\Bigl(\tfrac{\omega_{\rm q}^2(0)}{\omega_{\rm q}^2(\Phi)} + 1\Bigr) \Bigl[\sqrt{\tfrac{2 \Delta}{\delta \Delta - \hbar\omega_{\rm q}(\Phi)}} + \sqrt{\tfrac{2 \Delta}{\delta \Delta + \hbar\omega_{\rm q}(\Phi)}}\,\Bigr] - \tfrac{1}{4\pi}\Bigl(\tfrac{\omega_{\rm q}^2(0)}{\omega_{\rm q}^2(\Phi)} - 1\Bigr) \sqrt{\tfrac{2 \Delta}{\delta \Delta}}\,\Bigr] x_{{\rm qp}, L} - \tfrac{1}{4}x_{{\rm qp}, H},
\end{equation}
where $\Phi$ is the flux bias controlling the qubit frequency.
To cast the result in this form, we used the relation $\omega_{\rm q}^2(\Phi) = \sum_j \cos \varphi_j / (L_j C)$.
We obtain Eq.~\eqref{eq:a_coefficient} by taking $\Phi \approx 0$ and assuming that the QP density on the high-gap side is negligible, $x_{{\rm qp}, H} \approx 0$.
\end{widetext}

Equation \eqref{eq:freq_shift_final} is valid under the assumption that the characteristic QP energies are small, $E \ll \delta \Delta - \hbar \omega_{\rm q}$. Generalization of that equation to arbitrary energy distribution of QPs is provided by the combination of Eqs.~\eqref{eq:shift_vs_ImY}, \eqref{eq:ind_and_dyn}, \eqref{eq:ind_qp}, \eqref{eq:I_qp_n}, \eqref{eq:I_qp_Delta}, and \eqref{eq:ImY_dyn_general}.

\section{Additional information on the QEC experiments}\label{appendix:qec}
\subsection{Details of the interleaved experiment analysis \label{appendix:qec_details}}
The analysis procedure to detect error bursts in the interleaved experiment is close to that described in Appendix~\ref{appendix:analysis}. 
The interleaved experiment includes three types of measurements: Ramsey measurements \bm{$\mathrm{R}$}, energy relaxation measurements \bm{$\mathrm{T_1}$}, and repetition code measurements \bm{$\mathrm{M}$}.
The measurements are designed such that, at each time sample, if no errors occurred, the qubits performing \bm{$\mathrm{R}$} end up in $\ket{0}$, the qubits performing \bm{$\mathrm{T_1}$} end up in $\ket{1}$, and the repetition code measure qubits \bm{$\mathrm{M}$} return the same state as in the previous time sample, i.e., detection result 0. 
We denote the number of qubits in a wrong state at each time sample as $\Sigma(t)$, $\Sigma_{T_1}(t)$, and $\Sigma_{\rm QEC}(t)$ for $\mathbf{R}$, $\mathbf{T_1}$, and $\mathbf{M}$, respectively.

We find error bursts by focusing on $\Sigma(t)$ and $\Sigma_{\rm QEC}(t)$ time series. To suppress the noise in the data, we first compute a moving average of $\Sigma(t)$ and $\Sigma_{\rm QEC}(t)$ over a $\pm 25~\text{ms}$ window about $t$ and subtract it from the respective signals; we denote the results of the subtraction by $\delta\Sigma(t)$ and $\delta\Sigma_{\rm QEC}(t)$. 
Then, we apply a matched filter to each of these two traces, with an exponential template \eqref{eq:app_template}. 
We pick $\tau_\text{MF} = 250~\mu\text{s}$.
We note that this value is shorter than the one used in Appendix~\ref{appendix:analysis}.
The interleaved experiment has a lower sensitivity than the experiments of Secs.~\ref{sec:t2_t2e_t1} and \ref{sec:tomo}, resulting in a shorter perceived duration of the bursts. 
We identify bursts when either of the two filtered signals reaches a value higher than 1.
The choice of the threshold is motivated by a desire to avoid false positives due to fluctuations at a single qubit level.
For each identified impact, we record the match-filtered values ${\rm MF}[\Sigma(t)]$ and ${\rm MF}[\Sigma_{\rm QEC}(t)]$ at the time of the impact. These values are shown in Fig.~\ref{fig:fig5}(d).

\subsubsection*{Filtering out the false positive events}
\begin{figure}[t]
  \begin{center}
    \includegraphics[scale = 1]{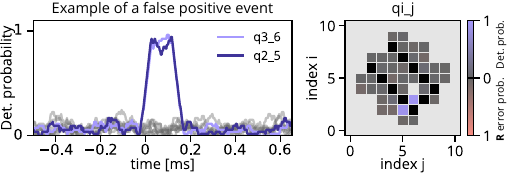}
    \caption{{\bf Example of a false positive event.} 
    (Left) Detection probability as a function of time for different measure qubits [we obtain the probability by counting the number of detection events in a $50\,{\rm \mu s}$ window]. Two of the measure qubits abruptly start to signal an error with probability close to $1$ for $\approx 150\,{\rm \mu s}$ [data for other qubits is shown with gray curves].
    (Right) The two signalling qubits are located at the adjacent sites. 
    The plot shows the detection probability on measure qubits, and Ramsey error probability on monitor qubits at averaged over the first $50{\rm \mu s}$ of the burst starting at $t = 0,{\rm \mu s}$.
    The detection/error probability remains close to zero on all qubits except for the two signalling ones.}
    \label{fig:fig_app_qec_false_positive}
  \end{center}
\end{figure}

Visual inspection of time sections identified as impact events by the matched filter reveals a number of cases within the repetition code time traces with a distinct temporal profile. 
Instead of a burst-like fast rise that is followed by the gradual decay of errors, these events feature the detection probabilities of two neighboring measure qubits abruptly inverting to values close to 1, staying at this level for between $50~\mu\text{s}$ to $250~\mu\text{s}$, and then sharply returning to baseline, see Fig.~\ref{fig:fig_app_qec_false_positive}. 
We posit that these events do not result from radiation impacts.
Rather, they occur when a data qubit situated between the two measure qubits becomes transiently unresponsive, potentially due to interactions with a two-level system or leakage out of the computation space.
The fact that the detection probability becomes close to $1$ is specific to such events: for coherent errors (which is the dominant source of errors during the radiation impacts), the detection probability is bounded by $1/2$ [see Fig.~\ref{fig:fig6}].

To distinguish and exclude these box-like events from gradually decaying impact-induced error bursts, we apply an additional filtering procedure to all time sections flagged as impact events by the initial matched filter analysis.
First, we convolve error traces from individual qubits with a $50\,\mu\text{s}$ box function, and check if the convolved signal exceeds $50\%$ error rate for more than 20 time steps ($\approx 20\,{\rm \mu s}$). 
If there are qubits like that, then for them, we run a procedure aimed at classifying if the time trace is box-like or gradually decaying.
Specifically, we count the number $n$ of time samples above threshold $p_{\rm th}$ for a variable threshold.
We expect that for a box-like function, the number $n(p_{\rm th})$ would stay relatively constant with $p_{\rm th}$, while for the decaying function $n(p_{\rm th})$ would decrease with $p_{\rm th}$.
We choose thresholds to range from $3\sigma$ above background to $1$ in steps of $0.1$ [$\sigma$ is the standard deviation of the signal]. 
We fit the dependence of $n(p_{\rm th}) / n(3\sigma)$ on $p_{\rm th}$ to a linear function; if the slope is small ($ < 2$), then we classify the error signature as box-like.
Any time section containing at least one qubit exhibiting such a box-like signature is removed from our list of impact events.

\begin{figure}[t]
  \begin{center}
    \includegraphics[scale = 1]{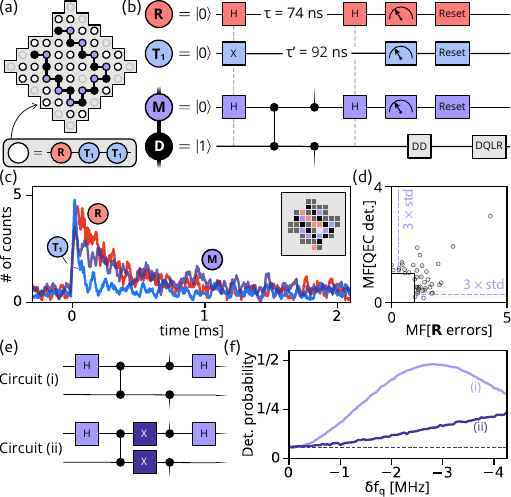}
    \caption{{\bf Results for the Z-basis repetition code experiments.} 
    {\bf (a)} The qubit layout is similar to that in the X-basis repetition code experiments, see Sec.~\ref{sec:interleaved}. {\bf (b)} The repetition code and coherence characterization circuits. The two qubit gates used in the repetition code are CZ gates. {\bf (c)}~Example of an error burst in the Z-basis repetition code detections (purple), \textbf{R} measurements (red), and $\mathbf{T_1}$ measurements (blue). The data is processes in the same way as in Fig.~\ref{fig:fig5}.
    Inset: spatial distribution of QEC detections and \textbf{R} errors at the start of the burst. 
    {\bf (d)} Correlation between repetition code detection bursts and error bursts in \textbf{R}. The plot shows the values of the matched filter (MF) applied separately to QEC and \textbf{R} data, at the time of the impact. Purple dashed lines and solid black lines have the same meaning as in Fig.~\ref{fig:fig5}.
   {\bf (e)} In the frequency shift injection experiment, we use two circuits that differ by the presence of an echo pulse between the two CZ gates. 
   {\bf (f)} Dependence of the average detection probability in the Z-basis repetition code on $\delta f_{\rm q}$.}
    \label{fig:fig_app_qec_z}
  \end{center}
\end{figure}

\subsection{Bit-flip repetition code experiments}

In addition to the X-basis repetition code experiments, we perform a set of experiments with the Z-basis repetition code.
This is a version of the repetition code that is designed to correct for bit-flip errors on data qubits.

We start with an interleaved monitoring experiment; the experiment is similar to the one in Sec.~\ref{sec:interleaved}, up to a basis change of the repetition code [cf.~Fig.~\ref{fig:fig5}(a,b) and Fig.~\ref{fig:fig_app_qec_z}(a,b)].
We observe that the temporal profile of repetition code bursts is close to the profile of {\it phase} error bursts on monitor qubits, see Fig.~\ref{fig:fig_app_qec_z}(c).
We further illustrate the correlation between the two types of bursts in Fig.~\ref{fig:fig_app_qec_z}(d).

The observed correlation between QEC detections and phase errors may come as a surprise for the Z-basis repetition code.
Naively, one would expect that this code is only sensitive to bit-flip errors, and not to phase errors.
While our circuit implementation of the code is indeed insensitive to phase errors on {\it data} qubits, it is not insensitive to phase errors on {\it measure} qubits.
We assert that it is these errors that are primarily responsible for the observed detection bursts. The measure qubits are susceptible to phase errors in a part of the circuit between the two Hadamards.
We can reduce the susceptibility by including an echo pulse in that part, see Fig.~\ref{fig:fig_app_qec_z}(e). 
The effectiveness of this strategy is illustrated by the controlled frequency shift injection experiment, see Fig.~\ref{fig:fig_app_qec_z}(f) [the experiment is similar to that in Sec.~\ref{sec:injection}].
We observe that addition of the echo pulse strongly reduces the effect of frequency shifts on the detection probability.

\subsection{Modeling of frequency shift-induced QEC detections}\label{appendix:inj_modeling}

To understand mechanisms by which frequency shifts cause QEC detections, we numerically simulate the X-basis repetition code circuits of Sec.~\ref{sec:qec}. 
We show that the results of the frequency shift injection experiment [see Sec.~\ref{sec:injection} and Fig.~\ref{fig:fig6}(a)--(c)] are consistent with a model which includes shift-induced phase and gate errors.

Let us explain the model by first focusing on two-qubit gate operations, that is, CZ and data qubit leakage removal (DQLR) gates \footnote{We refer to DQLR gate \cite{miao_2023} as to a two-qubit operation because in it, the leakage state of a data qubit is transferred to a neighbouring measure qubit.}.
Uniform frequency shifts used in the injection experiment commute through these gates, and so do not affect the gate fidelities per se.
However, the gates have a finite duration $t_{\rm gate}$; in this duration, an undesired phase accumulation occurs.
We model the phase accumulation by unitaries $U_\varphi = \exp(i\varphi \sigma_z / 2)$ with phase $\varphi = 2\pi \delta f_{\rm q} t_{\rm gate}$ \footnote{There is a subtlety in evaluating $\varphi$ for the injection experiment coming from the fact that we implement frequency shifts as fixed flux offsets.
Indeed, during the two-qubit gates, we move qubits in frequency \cite{foxen2020}; this changes the qubits sensitivity to flux offsets.
As a result, the frequency shift ``perceived'' by the qubit during the two-qubit gate differs from its nominal value by, on average, $28\%$. We include this effect in our modeling by adjusting $\varphi$ accordingly.}. 

Frequency shifts also introduce errors of single-qubit microwave gates, namely, Hadamard and $\pi$ pulses. Indeed, both gates rely on the application of a drive resonant with the qubit $|0\rangle \leftrightarrow |1\rangle$ transition.
The frequency shift brings the drive out of the resonance, and thus degrades the gate fidelity. 
The degradation may be described by a single-qubit Hamiltonian
\begin{align}  \label{eq:QubitHamiltonian}
	H_{\rm 1q} &= \frac{2\pi \delta f_{\rm q}}{2} \sigma_z + \frac{\Omega(t)}{2} \sigma_y,  
\end{align}
where $\Omega(t)$ is the drive amplitude.
We neglect the $|2\rangle, |3\rangle, \dots$ states of the qubit in Eq.~\eqref{eq:QubitHamiltonian}, and model the gates as simple cosine-envelope pulses with no DRAG corrections \cite{motzoi_2009}. To obtain the unitaries describing the faulty gates, we numerically evaluate $U_{\rm 1q} = {\cal T} \exp[-i\int dt H_{\rm 1q}(t)]$, where ${\cal T}$ denotes the time ordering.

We combine these error mechanisms into a single numerical model of the full repetition code experiment. 
We use the model to find how the probability of the detection events depends on the frequency shift $\delta f_{\rm q}$.
Detection events correspond to instances in the measurement record in which two subsequent measurement outcomes on a given qubit differ.
We find the detection event probabilities by sampling quantum trajectories of the system's wavefunction across many cycles of error correction.
We consider cases of 3 and 5 data qubits and verify that results are independent of the system size. 
We note that in the experiment, the detection event probability is nonzero at $\delta f_{\rm q} = 0$, see Fig.~\ref{fig:fig6}.
The respective detection events result from background sources of errors unrelated to frequency shifts.
We account for the background heuristically by adding a random bit flip probability to the measurement record before counting detections, with probability ensuring a correct number of detections at zero shift.
The described model has no free parameters.

\begin{figure}[t]
  \begin{center}
    \includegraphics[scale = 1]{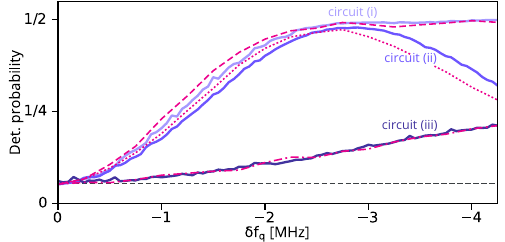}
    \caption{\textbf{Predicted QEC detection probabilities for the frequency shift injection experiments.} Magenta curves show the prediction of our parameter free model; dashed, dotted, and dash-dotted lines correspond to circuits (i), (ii), and (iii), respectively. Solid lines reproduce the experimental data of Fig.~\ref{fig:fig6}(c).} 
    \label{fig:app_qec_simulation}
  \end{center}
\end{figure}

The results of simulation of circuits (i)--(iii) are shown in Fig.~\ref{fig:app_qec_simulation} alongside the experimental data of Fig.~\ref{fig:fig6}.
There is a good agreement between the data and the simulation across the considered range of frequnecy shifts. 
In circuits (i) and (ii), the detection probability at small frequency shifts ($\lesssim$ 1 MHz) is dominated by the phase accumulation during the CZ gates. At the largest shifts, phase errors during DQLR also contribute substantially to circuit (i); however, these errors are echoed away in circuit (ii). The downturn in detection probability at $\delta f_{\rm q} \approx 2.8$ MHz in circuit (ii) occurs because the total accumulated phase on the measure qubits between the two Hadamard gates begins to exceed $\pi/2$, at which point the effective bit-flip  probability $p_{\rm meas}$ starts to exceed $1/2$. The detection probability is given by $p_{\rm det} = 2p_{\rm meas} (1 - p_{\rm meas})$; it decreases with increasing $p_{\rm meas}$ for $p_{\rm meas} > 1/2$. 
We can estimate the frequency at which this downturn occurs as follows. Phase accumulation between the two Hadamard gates has contributions from the CZ gate durations, and from the Hadamard gates themselves. By solving Eq.~\eqref{eq:QubitHamiltonian} in the interaction picture to first order in $\delta f_{\rm q}$, we find that the phase contributed by the Hadamard gates is approximately $0.58\times t_\text{Hadamard} \delta f_{\rm q}$ per gate. The total phase accumulated
is thus $\varphi \approx 2\pi \delta f_{\rm q}(2 t_{\rm CZ} + 2 \times 0.58 \times t_\text{Hadamard})$. We find that $\varphi = \pi/2$ at $\delta f_{\rm q} \approx 2.6$ MHz, close to the observed inflection point.

In circuit (iii), all free-evolution periods are coherently cancelled. The remaining error arises from errors of single-qubit gates. 
To quantify the gate errors due to the frequency shifts, we evaluate the gate fidelity.
Solving Eq.~\eqref{eq:QubitHamiltonian} to {second} order in $\delta f_{\rm q}$ in the interaction picture, we find for the fidelity of a $\pi$ pulse:
\begin{align}
	\text{Fid}[U_\pi] = 1 - \frac{(0.39 (2\pi \delta f_{\rm q}) t_\text{pulse})^2}{6}
\end{align}
For example, at $\delta f_{\rm q} = 3$ MHz, this formula gives an infidelity of approximately 0.6\%.
We note that circuit~(iii) contains a number of single-qubit gates whose effects add up coherently. 
This is why the detection probability at $\delta f_{\rm q} = 3$ MHz is an order of magnitude larger than the infidelity of a single gate.



\bibliography{bib.bib}

\end{document}


\widetext

\title{
Supplemental Materials for\\ ``Correlated Error Bursts in a Gap-Engineered Superconducting Qubit Array''
}

\author{Vladislav D. Kurilovich}
\equal
\author{Gabrielle Roberts}
\equal
\author{Leigh S.~Martin}
\author{Matt McEwen}
\author{Alec Eickbusch}
\author{Lara Faoro}
\author{Lev B.~Ioffe}
\author{Juan Atalaya}
\author{Alexander Bilmes}
\author{John Mark Kreikebaum}
\author{Andreas Bengtsson}
\author{Paul Klimov}
\author{Matthew Neeley}
\author{Wojciech Mruczkiewicz}
\author{Kevin Miao}
\author{Igor L.~Aleiner}
\author{Julian Kelly}
\author{Yu Chen}
\author{Kevin Satzinger}
\author{Alex Opremcak}
\equal
\google

\maketitle
\newpage


\begin{figure*}[t]
\begin{center}
\includegraphics[width=1\textwidth]{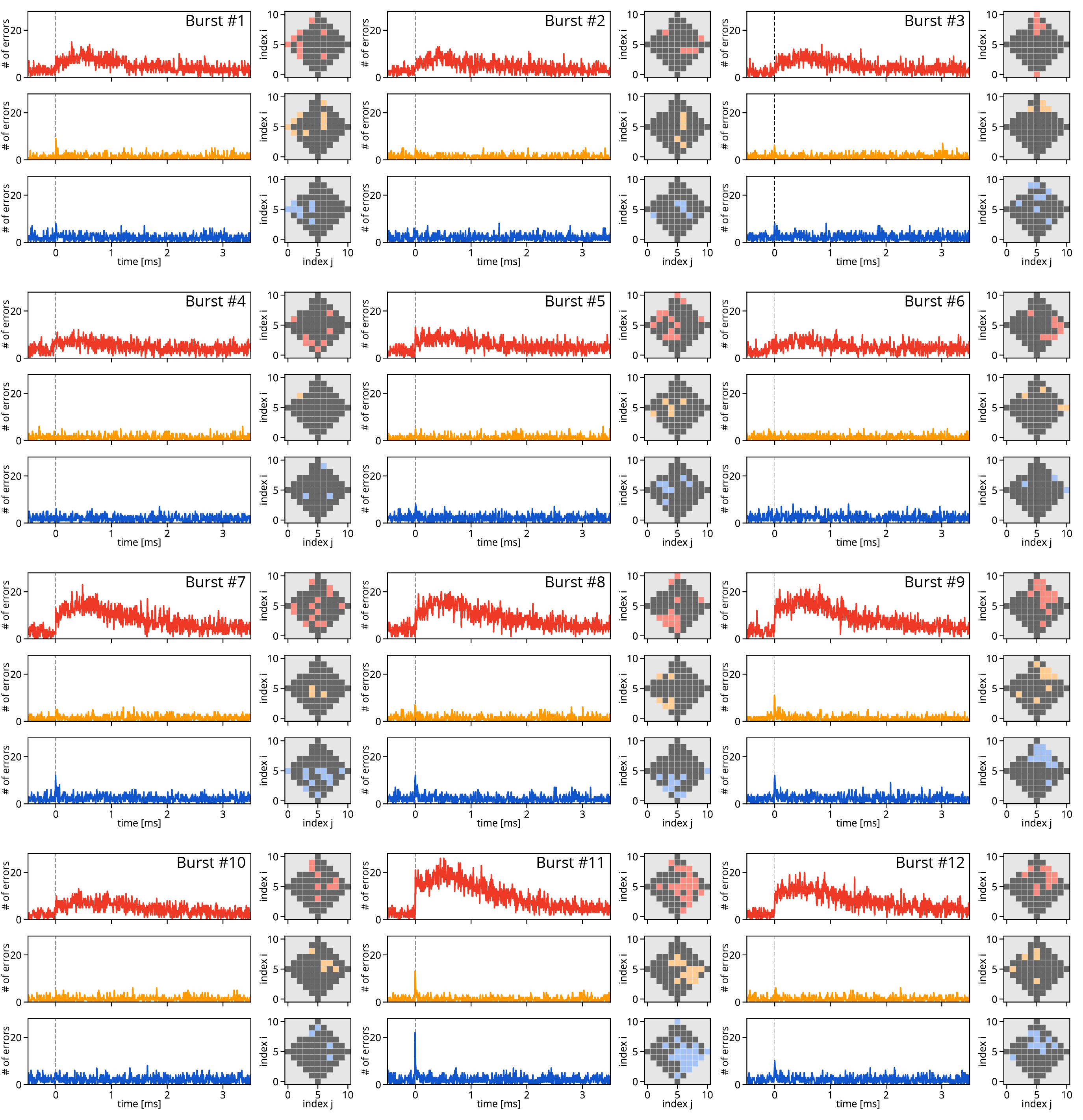}
\caption{{\bf Additional examples of error bursts in R, E, and $\mathbf{T_1}$ measurements}. A sequence of twelve error bursts that were detected by our procedure one after the other. Real-space error distributions correspond to a moment depicted in the plots by a vertical dashed line.}
\label{fig:fig_supp_more_data_t2t2et1}
\end{center}
\end{figure*}

\begin{figure*}[t]
\begin{center}
\includegraphics[width=\textwidth]{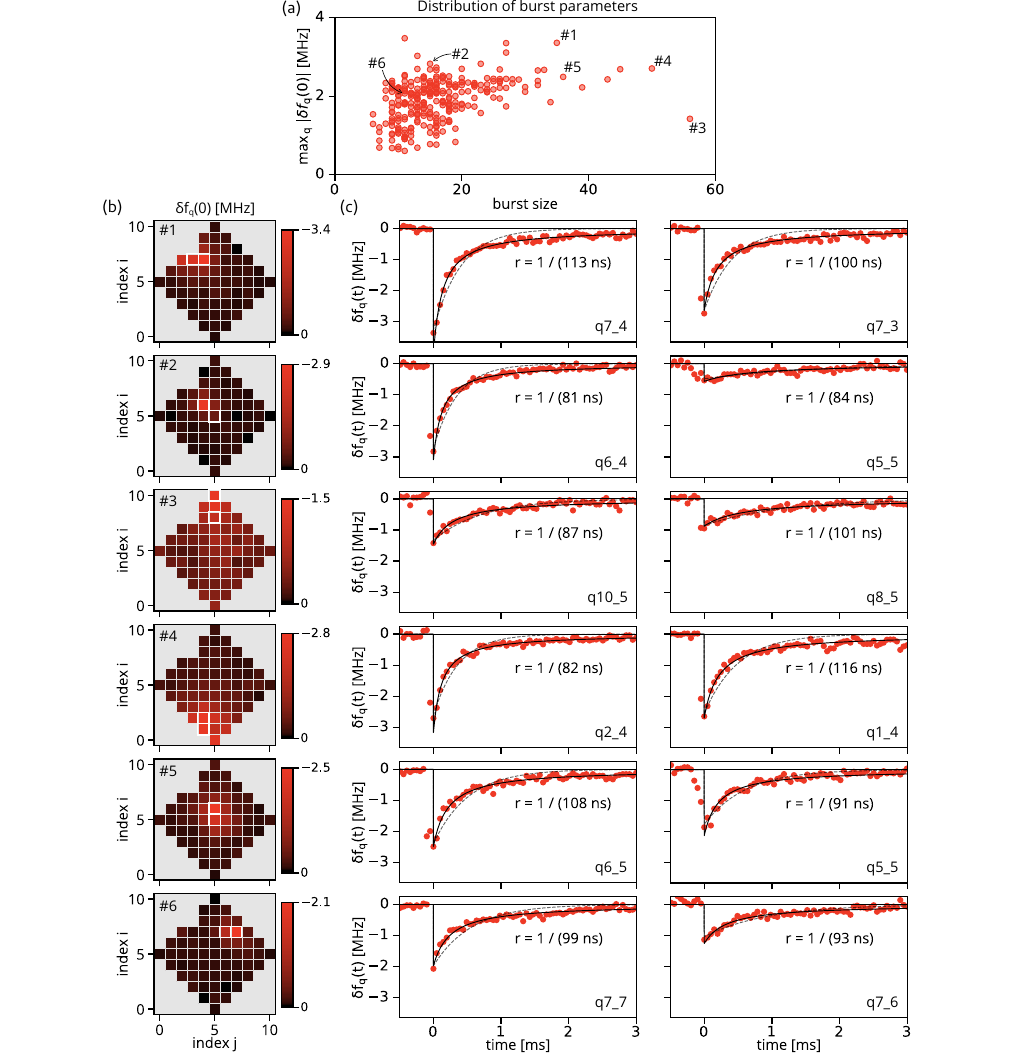}
\caption{{\bf Additional data on qubit frequency shifts $\bm{\delta f_{\rm q}(t)}$ during the bursts.} \textbf{(a)} We present data on six bursts of different sizes labelled $\#1-\#6$ [the distribution in this panel is identical to that in Fig.~5 of the main text]. \textbf{(b)} Position-dependence of the frequency shifts at the beginning of the burst for the six selected bursts. \textbf{(c)} Time-dependence of the shifts on two qubits in each burst. 
In all cases, $\delta f_{\rm q}(t)$ is well-approximated by Eq.~(3) of the main text (solid black lines). Dashed lines are provided for comparison, and show an exponential fit to the data.}
\label{fig:fig_supp_more_data}
\end{center}
\end{figure*}

\begin{figure*}[t]
\begin{center}
\includegraphics[width=\textwidth]{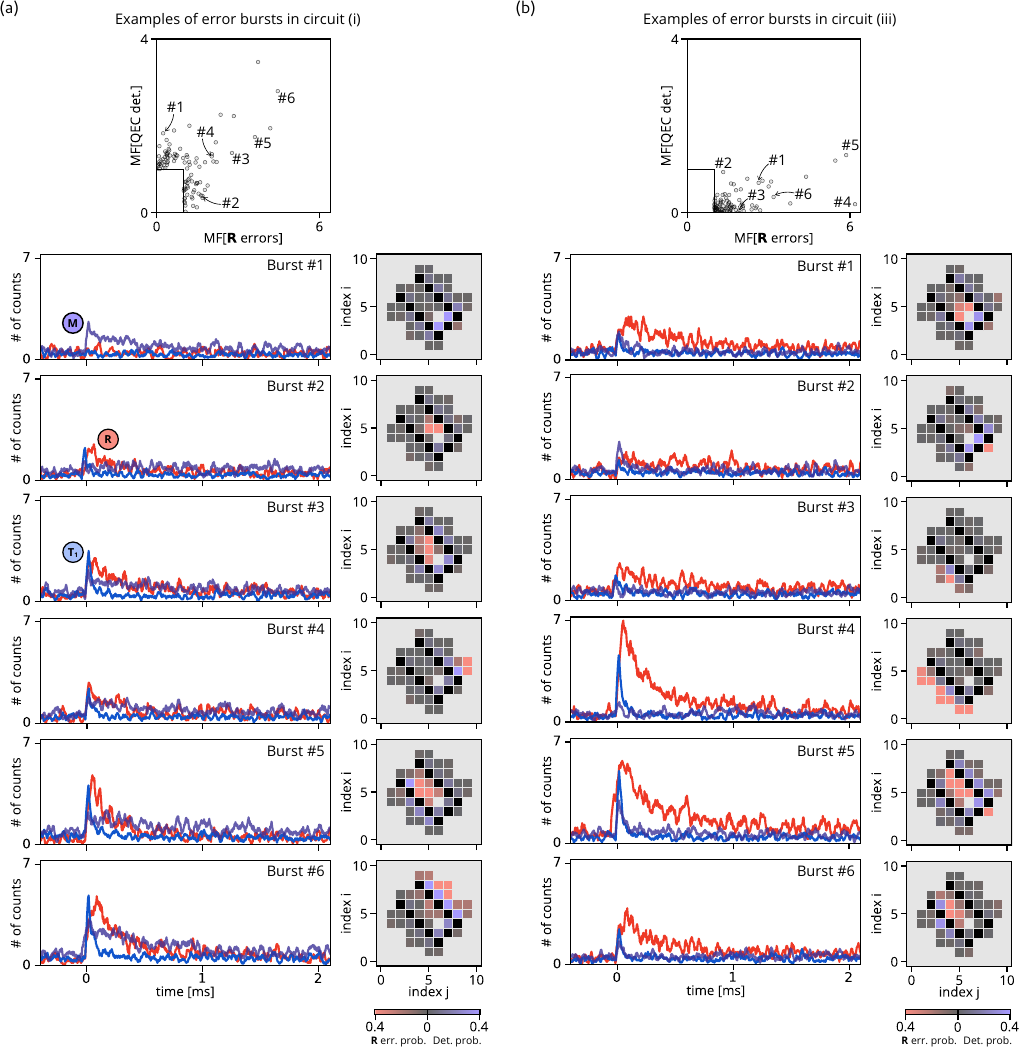}
\caption{{\bf Additional examples of error bursts in the interleaved experiment with the X-basis repetition code.} \textbf{(a)} Experiments with circuit (i) [circuits are shown in Fig.~7 of the main text]. We present data on six bursts of different sizes labelled $\#1-\#6$ (the burst distribution in the upper panel is identical to that in Fig.~6). Purple, red, and blue curves correspond to repetition code detections, Ramsey errors, and $T_1$ errors, respectively. Spatial distributions show probabilities of Ramsey errors on monitor qubits and probabilities of detection events on QEC measure qubits. We find probabilities by counting the total number of errors/detections the qubit experienced in $50\,{\rm \mu s}$ time-window around the peak of the burst. \textbf{(b)}~Same for circuit (iii).}
\label{fig:fig_supp_qec_examples}
\end{center}
\end{figure*}

\begin{figure*}[t]
\begin{center}
\includegraphics[width=\textwidth]{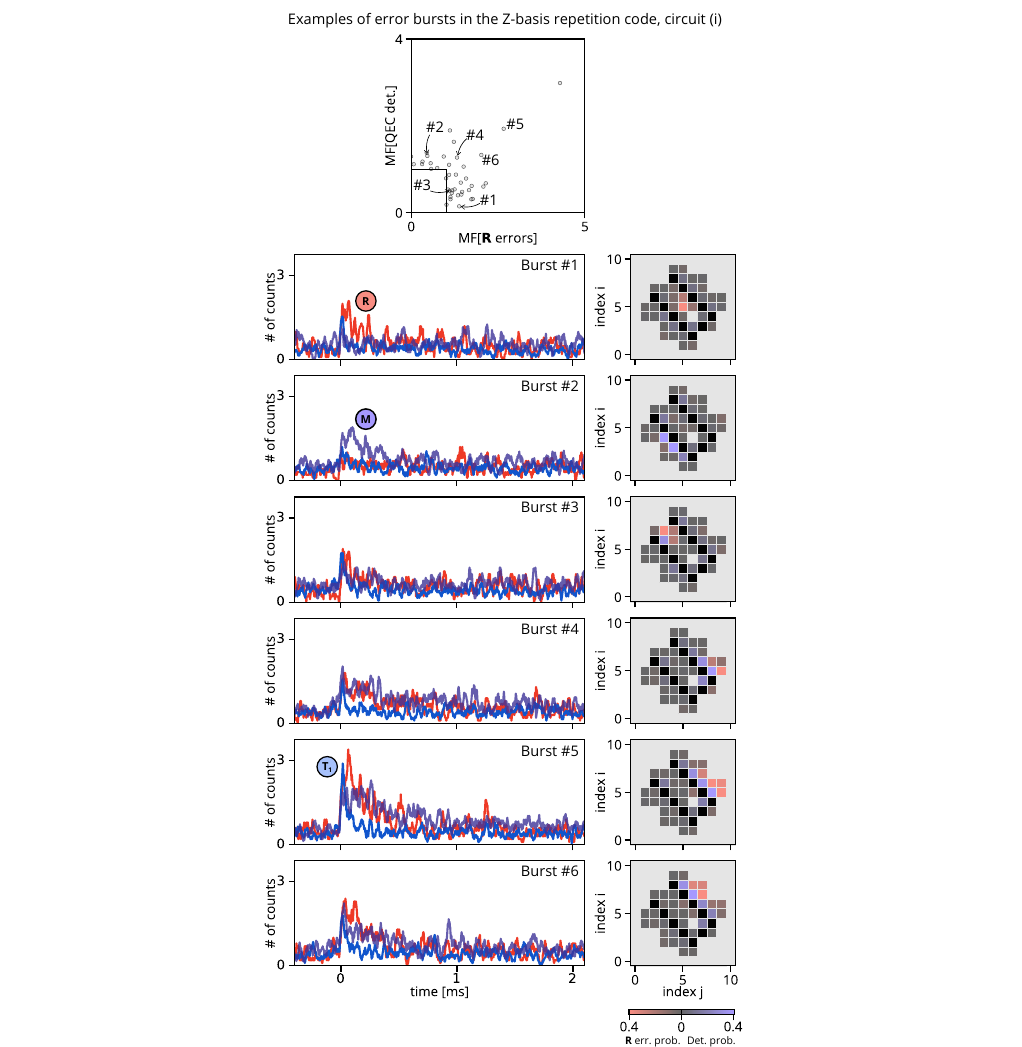}
\caption{{\bf Additional examples of error bursts in the interleaved experiment with the Z-basis repetition code.} Notations are similar to Fig.~\ref{fig:fig_supp_qec_examples}.}
\label{fig:fig_supp_qec_examples_Z}
\end{center}
\end{figure*}